\documentclass{ws-jai}
\usepackage{tabularx}
\usepackage[flushleft]{threeparttable}
\usepackage{tabularx}
    \newcolumntype{L}{>{\raggedright\arraybackslash}X}
\usepackage[normalem]{ulem}
\usepackage{color}

\usepackage{hyperref}
\catchline{}{}{}{}{} 
\begin{document}

\markboth{Podczerwinski and Timbie}{Design of an Ultra-Wideband Antenna Feed and Reflector for use in Hydrogen Intensity Mapping Interferometers}

\title{Design of an Ultra-Wideband Antenna Feed and Reflector for use in Hydrogen Intensity Mapping Interferometers}

\author{John Podczerwinski$^{1}$ and Peter Timbie$^{1}$}

\address{
$^{1}$Department of Physics, University of Wisconsin Madison, Madison, WI 53703, USA
}

\maketitle


\begin{history}
\received{(to be inserted by publisher)};
\revised{(to be inserted by publisher)};
\accepted{(to be inserted by publisher)};
\end{history}

\begin{abstract}
This paper describes the design of a 5.5:1 bandwidth feed antenna and reflector system, intended for use in hydrogen intensity mapping experiments. The system is optimized to reduce systematic effects that can arise in these experiments from scattering within the feed/reflector and cross-coupling between antennas. The proposed feed is an ultra wideband Vivaldi style design and was optimized to have a smooth frequency response, high gain, and minimal shadowing of the reflector dish. This feed can optionally include absorptive elements which reduce systematics but degrade sensitivity. The proposed reflector is a deep parabolic dish with $f/d = 0.216$ along with an elliptical collar to provide additional shielding. The procedure for optimizing these design choices is described.
\end{abstract}

\section{Introduction}
\noindent This paper describes the design of  a 5.5:1 bandwidth feed antenna and reflector intended for applications in radio astronomy, particularly for the proposed Packed Ultra-wideband Mapping Array (PUMA) telescope \cite{cosmic_visions}. This feed will be referred to as the ``proposed feed" in the paper. The proposed PUMA telescope is a radio-interferometer consisting of 32K tightly packed antennas. Each antenna will consist of a feed suspended over a 6\,m diameter parabolic reflector. These antennas are planned to have a bandwidth from 200 to 1100\,MHz, with extra emphasis placed on the performance from 200 to 475\,MHz.    

The paper also describes a version of the feed, referred to as the ``minimum systematics feed". This feed is similar to the proposed PUMA feed but includes absorptive elements that decrease systematic effects but increase the noise temperature of the antenna system. It will be argued that this design may be useful for telescopes operating at lower frequencies than PUMA.

A major goal of observational cosmology is to measure the statistical properties of the spatial distribution of matter in the Universe. Such measurements are of interest as they could tell us about dark energy, inflationary physics, the growth of structure, early star formation and more. The distribution of matter in the Universe is traced by neutral hydrogen (HI). Thus, observations of the 21\,cm line from HI can probe the matter distribution (up to a bias term). Using 21\,cm emission or absorption to measure these density fluctuations without resolving individual galaxies is a technique referred to as 21\,cm (or HI) intensity mapping \cite{Chang2008}. Because of the expansion of the Universe, the 21\,cm line will be redshifted by an amount that scales with the distance between the emitting region and Earth. For this reason, different epochs of the Universe can be probed based on the range of observing frequencies chosen. The information that can be extracted from the distribution varies with cosmological redshift, denoted by $z$. Measurements of the distribution at redshifts $z \leq 6$ could probe the growth of structure, the physics of dark energy, and inflationary physics. Measurements during the Epoch of Reionization (EOR) ($6 < z < 20$) would tell us about the formation of the first stars. Lastly, measurements during the cosmic dark ages ($20 < z < 1100$), an epoch preceding the first stars and relatively unaffected by astrophysical phenomena, would probe the physics of inflation. The 200 to 1100\,MHz bandwidth of PUMA is chosen to capture redshifts $0.3 \leq z \leq 6$.  

For reasons of sensitivity and cost, radio-interferometers are typically used for HI intensity mapping. In interferometry, an array consisting of many antenna elements is used. The signals from pairs of antennas are correlated to form ``visibilities". In the absence of mutual coupling, correlating these signals eliminates the offset from system noise one sees in single dish telescopes. This in turn relaxes requirements for gain stability. Moreover, combining signals from many pairs of antennas allows the same angular scale to be measured many times, providing a distinct advantage over single dish experiments. 

Such interferometers are optimized to measure the spatial power spectrum $P(k)$ of the 21\,cm line. Cosmologists are particularly interested in large spatial scales, corresponding to small values of the wavenumber $k$. This interest in large spatial scales leads to interferometers with tightly packed antenna elements. In the case of PUMA, at the longest wavelength, dishes are planned to be about $4\lambda$ wavelengths in diameter with centers located about $5\lambda$ from each other. This spacing is small enough to probe large spatial scales at all wavelengths, while also being large enough to allow dishes to be repointed to latitudes away from zenith without beam blockage. Moreover, the HI signal being measured is small, with a brightness temperature of roughly 1\,{mK}, while system temperatures are the order of $\sim25-100$\,K in the post EOR universe and up to $\sim1000$\,K in the case of EOR measurements. The small magnitude of the signal being measured motivates interferometers with large collecting areas. These requirements lead to `large N, small D' designs, which consist of large numbers of small diameter antennas. Although this paper focuses on the proposed PUMA telescope, other designs exist as well. Table \ref{table:telescopes} provides summaries of several interferometers intended for measuring the EOR and post EOR Universe.

\begin{wstable}[h] 
\caption{Summary of a few interferometers intended for 21\,cm intensity mapping. Fiducial values for PUMA's aperture efficiency and amplifier temperature are used.}
\begin{tabular}{@{}p{3cm}p{4cm}p{2.2cm}p{3cm}p{2cm}@{}} \toprule
 & \textbf{Antenna} & \textbf{Bandwidth} & \textbf{Array Geometry} &
\textbf{Effective Aperture per Element}\\
 \colrule
\textbf{PUMA}\cite{cosmic_visions} & 6\,m diameter parabolic dish, feed on-axis & \hphantom{0}200-1100\,MHz $6 \geq z \geq 0.3$ (Post EOR) & 32K tightly packed antennas. & $\hphantom{0}\approx 20\mathrm{m}^{2}$  \\ \colrule
\textbf{HIRAX}\cite{hirax} & 6\,m diameter parabolic dish, feed on-axis & \hphantom{0}400-800\,MHz $2.55 \geq z \geq 0.78$ (Post EOR)& 1024 elements in a $32 \times 32$ tightly packed array. & \hphantom{0}$15.5\mathrm{m}^{2}$ at 600\,MHz \\ \colrule 
\textbf{CHORD}\cite{CHORD} & 6\,m diameter parabolic dish, feed on-axis. Also includes 90\,m long 10\,m wide cylindrical reflectors with a focal ratio of $0.25$. & \hphantom{0}300-1500\,MHz $3.7 \geq z \geq 0$ (Post EOR) & 512 tightly packed dishes at the central station. Two outrigger stations placed 1500\,km and 3000\,km away. These outrigger stations have 64 dishes and 1 cylinder each. \\
\colrule
\textbf{HERA}\cite{HERA} & 14\,m diameter parabolic dish, feed on-axis & 50-250\,MHz $27 \geq z \geq 4.7$ (EOR) & 350 dishes tightly packed in a hexagonal array. 32 outrigger antennas placed several hundred meters from the center of the array. & $\hphantom{0}\approx 100\mathrm{m}^{2}$ at 150\,MHz  
\\ \botrule
\end{tabular}
\label{table:telescopes}
\end{wstable}

    

In an HI intensity mapping instrument, candidate antennas must be designed to minimize systematic effects. The primary systematic effect is contamination from Galactic synchrotron radiation. The Galactic foregrounds are of concern as they are about 5 orders of magnitude brighter than the predicted HI signal. However, the foregrounds have a smooth spectrum, while the spectrum of the HI signal is quite `spikey', corresponding to clumps of hydrogen gas along the line of sight. Thus, one can confine the Galactic foregrounds into a small region of the delay domain by ensuring that the antennas have a smooth response as a function of frequency. In 21\,cm intensity mapping, delay is the Fourier dual to spectral frequency \cite{delay_spec}. Moreover, one needs to minimize mutual coupling between antennas, which becomes challenging due to the close element spacing required in HI interferometers.   Such mutual coupling can degrade delay spectrum  performance, introduce correlated noise, and make certain calibration methods more difficult. Designs that minimize these systematic effects can be seen in \citet{hera_viv} as well as \citet{hirax}. Of course, one also needs to consider the sensitivity of the instrument to the HI power spectrum. Sensitivity will be given less priority during optimization than the previously mentioned systematics effects, since sensitivity can be improved by increasing the number of elements or observing for longer. 

Lastly, polarization leakage also should be considered. The desired HI signal is unpolarized, while the foregrounds do have a polarized component. Leakage of polarized foregrounds into estimates of unpolarized emission can be caused by beam ellipticity and by cross-polarization in the beam. Such leakage is an additional systematic effect one could consider when designing an antenna for HI intensity mapping. This design was not optimized with polarization leakage in mind, but cross-polarization and leakage estimates are still provided in Section \ref{subsection:crosspol}.

This paper is organized as follows. Section \ref{section:Design Requirements} describes the performance requirements of the instrument and the motivation for the requirements. Section \ref{section: Design and Opt} describes the optimization process for the feed antenna along with the resulting design. In addition, it shows simulations of the S-parameters and beam patterns. Section \ref{section: sim perf} summarizes the performance of both the proposed and minimum systematics antenna systems. Finally, Section \ref{section: fab} describes the fabrication of a $4/13$ scale version of the minimum systematics feed and provides some measured results.

\section{Performance Requirements}
\label{section:Design Requirements}
The feed antennas and reflector were designed with HI intensity mapping instruments such as PUMA in mind. PUMA is a large, tightly packed antenna array. In instruments such as PUMA, mitigation of systematic effects is of primary importance. In addition, one must also consider the sensitivity of the instrument to the signal being measured.


In Sections \ref{subsection:smooth} and \ref{subsection:XC}, aspects of performance most crucial to minimizing systematic effects are considered. In Section \ref{subsection:sensitivity}, sensitivity of the instrument to the HI signal is considered.

\subsection{Spectral Smoothness and the Foregrounds} \label{subsection:smooth}
Due to their smooth frequency spectrum, the foregrounds encountered in instruments such as PUMA are naturally contained in a region of 2-dimensional k-space (with k components parallel to and perpendicular to the line of sight) referred to as ``the wedge". When taking measurements, the wedge is convolved with the response of the instrument, causing foregrounds to leak out into otherwise signal-dominated modes. In 21\,cm intensity mapping, it is the case that delay ($\tau$) is proportional to line of sight wavenumber $k_{||}$.  It is thus desirable for the antenna response in the delay domain be concentrated at 0 delay and decay rapidly at larger delays. 

In order to test the smoothness of the antenna's response, we will excite the antenna with a planewave coming from some direction $\hat{n}$. The voltage at the terminals of the antenna is then given by 
\begin{equation}
    v(\hat{n},\nu) = \mathbf{r}(\hat{n},\nu) \cdot \mathbf{E}(\hat{n},\nu).
\end{equation}
In this formula, $\mathbf{r}$ is referred to as the ``voltage beam" of the antenna. We then quantify the delay performance of the antenna is via the Fourier transform of the power kernel. The power kernel is defined as  
\begin{equation}
    R(\hat{n},\nu) = |\mathbf{r}(\hat{n},\nu)|^{2},
\end{equation}
In particular, we will characterize the delay response of the instrument using the delay spectrum of the power kernel at zenith. This is denoted via $\tilde{R}(\theta=0,\phi=0,\tau)$. 
The goal is to have $\tilde{R}(0,0,\tau)$ drop by 50\,dB as quickly as possible as a function of $\tau$. Refer to \citet{delay_spec} for more details. 

\subsection{Mutual Coupling} \label{subsection:XC}
Another concern in HI intensity mapping instruments is mutual coupling between the antennas in the array. It has been found that mutual coupling changes the beam patterns of antenna elements as described in \citet{Fagnoni} and \citet{kerns}. This effect on the beam patterns will likely degrade spectral smoothness \cite{Fagnoni}. Moreover, it causes beam patterns to vary from element to element. Such beam pattern variation is undesirable as it would prevent the use of redundant calibration techniques which have been proposed to overcome the computational challenge of calibrating large, redundant, arrays. In redundant calibration, one assumes that antennas in the array have identical beam patterns. This assumption would not be valid in the case of significant mutual coupling \cite{redundant}.

Mutual coupling can also cause receiver noise radiated from one antenna to be absorbed by a different antenna.  This coupling creates ``correlated noise", which causes non-zero visibilities. These correlated noise contributions can masquerade as sky signals \cite{kwak_2022}. In principle, this coupling varies slowly in time, but it sets a requirement on the gain stability of the instrument.

During the design process, mutual coupling was characterized by simulating the $S_{21}$ coupling between two adjacent antennas. The antennas are placed 8\,m apart and pointed at zenith. Such a setup can be thought of as the worst case scenario for coupling between two antennas in arrays such as PUMA. Simulations were performed with antennas co-located in each other's E-planes and H-planes. Using the model for coupled signals described in \citet{kerns}, one should find that the ``correlated noise" between two antennas follows Eq \ref{eq:tcoup}.  
\begin{equation} \label{eq:tcoup}
    T_{coupled} \propto |S_{21}|T_{ampl}.
\end{equation}
Since $T_{ampl}$ is typically 4 to 5 orders of magnitude larger than $T_{HI}$, the $S_{21}$ must be kept below $-100$\,dB to prevent correlated noise from overpowering the signal. It should be acknowledged that the model presented in Eq \ref{eq:tcoup} is an approximation.  A more sophisticated model of this phenomenon is presented in \citet{kwak_2022}.

 \subsection{Sensitivity} \label{subsection:sensitivity}
In addition to minimizing systematic effects, we require this antenna (and array) to provide enough sensitivity to detect the HI signal.  
In such an measurement, the contribution of system noise to the power spectrum goes like \cite{Liu_Shaw,Parsons_2014}

\begin{equation}\label{eqn:parsons}
    P_{N} \propto \frac{\Omega_{p}^{2}}{\Omega_{pp}}T_{sys}^{2},
\end{equation}
Let $P$ be the power pattern of the antenna. We define $\Omega_{p}$ and $\Omega_{pp}$ in terms of $P$ via
\begin{equation}\label{eqn:bull}
    \Omega_{p} \equiv \int d\Omega P(\hat{\mathbf{n}})
\end{equation}
 and
\begin{equation}
    \Omega_{pp} \equiv \int d\Omega P(\hat{\mathbf{n}})^{2}. 
\end{equation}
In these formulas, the system temperature $T_{sys}$ is estimated via \cite{cosmic_visions}
\begin{equation}\label{eq:tsys}
    T_{sys} \approx \frac{T_{ampl}}{\eta_{sp}\eta_{c}} + \frac{1-\eta_{sp}}{\eta_{sp}} T_{ground} + T_{sky}.
\end{equation}
In this formula, $T_{ampl}$ is the noise temperature of the amplifier and $\eta_{sp}$ is defined as  
the fraction of the antenna's beam that shines on the sky.  One can compute $\eta_{sp}$ using 
\begin{equation}
    \eta_{sp} = 4\pi \int_{0}^{\frac{\pi}{2}} \int_{0}^{2\pi} d\theta d\phi G(\theta,\phi).
\end{equation}
The optical efficiency $\eta_{c}$ is defined as 
\begin{equation}
    \eta_{c} = \eta_{S11} \eta_{rad},
\end{equation}
where $\eta_{S11}$ is the fraction transmitted through the feed antenna's input terminals and $\eta_{rad}$ is the radiation efficiency. 

For the nominal PUMA design \cite{cosmic_visions}, the proposed values of these parameters for achieving adequate sensitivity are: $\eta_{sp} = 0.9$, aperture efficiency $\eta_{A} = 0.7$, $\eta_{c} = 0.9$ and $T_{ampl} = 50$\,K. Thus, the designs proposed in this paper ought to provide comparable sensitivity to these fiducial parameters.

 The performance requirements for the antennas proposed in this paper are summarized in Table \ref{table:goals}.
\begin{wstable}[h]
    \caption{The primary performance requirements for the antenna described in this paper.}
    \begin{tabular}{@{}l p{10cm}@{}} \toprule  
       \textbf{Aspect of Performance} & \textbf{Goal} \\ \colrule   
       \textbf{Delay Response} & The antenna response at zenith decays by 50\,dB as quickly as possible in the delay domain. As a realistic goal, we wanted to keep the $50\,dB$ half width $\lesssim 0.10hMpc^{-1}$ when described in k-space. \\ \colrule
       
       \textbf{Mutual Coupling} & The $S_{21}$ of two antennas separated by 8\,m is at or below -100\,dB when the antennas are co-located in each other's E-plane or H-plane. \\ \colrule
       
       \textbf{Impedance Match} & The feed has $S_{11} \leq -10\,$dB. The final design presented will be relative to $200\,\Omega$. \\ \colrule
      
       \textbf{Power Spectrum Sensitivity} & Power spectrum sensitivity is comparable to that achieved by the fiducial PUMA parameters
       \cite{cosmic_visions}. This is achieved via a low $S_{11}$ and very high spillover efficiency ($\eta_{sp} \approx 0.99$).  \\ \colrule
       
       \textbf{Bandwidth} & All of these requirements are satisfied over the entire PUMA band,200 to 1100\,MHz. \\ \botrule
  
\end{tabular}
\label{table:goals}
\end{wstable}

\section{The Proposed Antenna and its Optimization}
\label{section: Design and Opt}
This section describes the electrical and mechanical design of the proposed system, and motivates the choices made to achieve the performance requirements. The proposed antenna system was designed with the electromagnetic simulation software package CST Microwave Studio \footnote{\url{http://www.cst.com}}. The work described in this paper primarily used CST's time domain solver. However, the frequency domain solver and integral equation solver were used at various times to double check results. All beam patterns presented in this work were computed using the ``open (add space)" boundary condition in CST.

\subsection{Dish Reflector} \label{subsection:dish}
In this paper, $6$\,m diameter parabolic dishes with $f/d = 0.216$ and $0.375$ are shown. These dishes are on-axis reflectors. Off-axis reflectors were also considered, but are not favored since they provide limited shielding against mutual coupling between antennas. The diameter of the dishes corresponds to $4\lambda$ at the longest wavelength. This small dish size was chosen since smaller dishes allow for a more tightly packed array, which in turn provides access to larger angular scales on the sky. This small dish size drove many design choices in the feed. In particular, it was one of the main reasons a Vivaldi-style feed was chosen, as opposed to a quad-ridge horn. The use of a horn antenna would cause too much shadowing when placed directly over a small dish.

A low $f/d$ ratio is favored in order to provide shielding for the feeds, which reduces mutual coupling. An example of the benefits of using a deeper dish can be seen in \citet{hirax}. In addition, outfitting the deep dish with a collar to provide extra shielding was considered. In this particular paper, a simple elliptical collar was used. Collars providing a smoother junction with the dish can be designed using the methods shown in \citet{Gupta}. Using a curved collar, whether elliptical or through Gupta's method, is desirable for two reasons. The first is that these methods minimize the discontinuity at the boundary between the parabola and the collar. This in turn minimizes diffracted fields emanating from that boundary. The second reason is that the curvature of the collar minimizes standing waves between the feed and collar. If this were not the case, then the presence of the collar would degrade the delay spectrum significantly.    

\begin{figure}[h]
    \centering
    \includegraphics[scale=0.21]{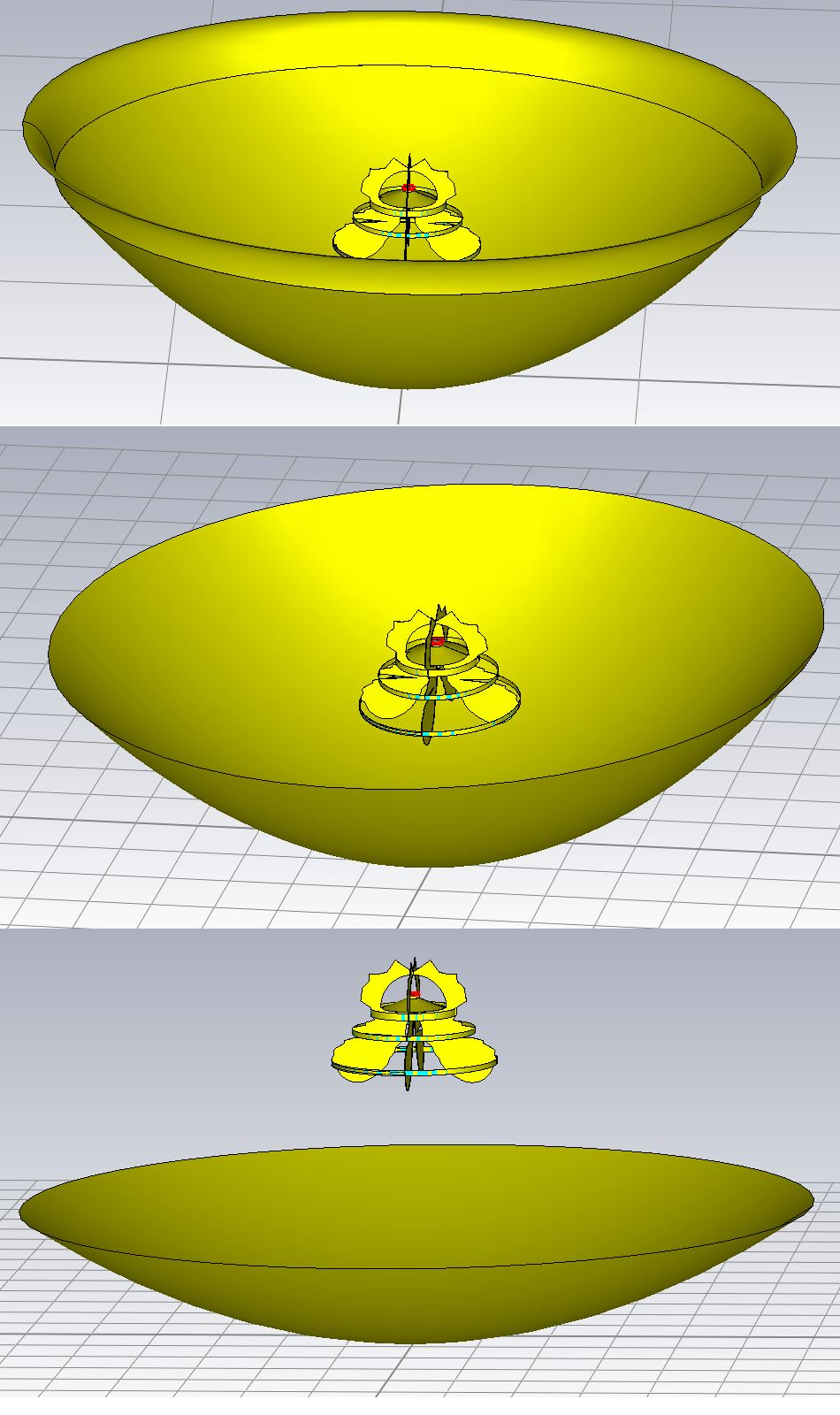}
    \caption{The three dishes considered in this paper. The top image shows the $f/d = 0.216$ dish with the elliptical collar included. The middle image shows the $f/d= 0.216$ dish without the collar. The bottom image shows the $f/d = 0.375$ dish.}
    \label{fig:dish}
\end{figure}

Some of the dishes considered can be seen in Figure \ref{fig:dish}. 
Decreasing $f/d$ increases reflections between the dish and feed, which in turn degrades delay spectrum performance. Fortunately, the reduced mutual coupling provided by a deep dish improves delay spectrum performance by cutting down on inter-dish reflections \cite{hirax}. It was found in the Saliwanchik \textit{et al} that these inter-dish reflections dominate over intra-dish reflections.  Thus, $f/d$ should be chosen as low as possible while still being mechanically feasible and providing adequate sensitivity. 

\begin{figure}
   \centering
    \includegraphics[scale=0.46]{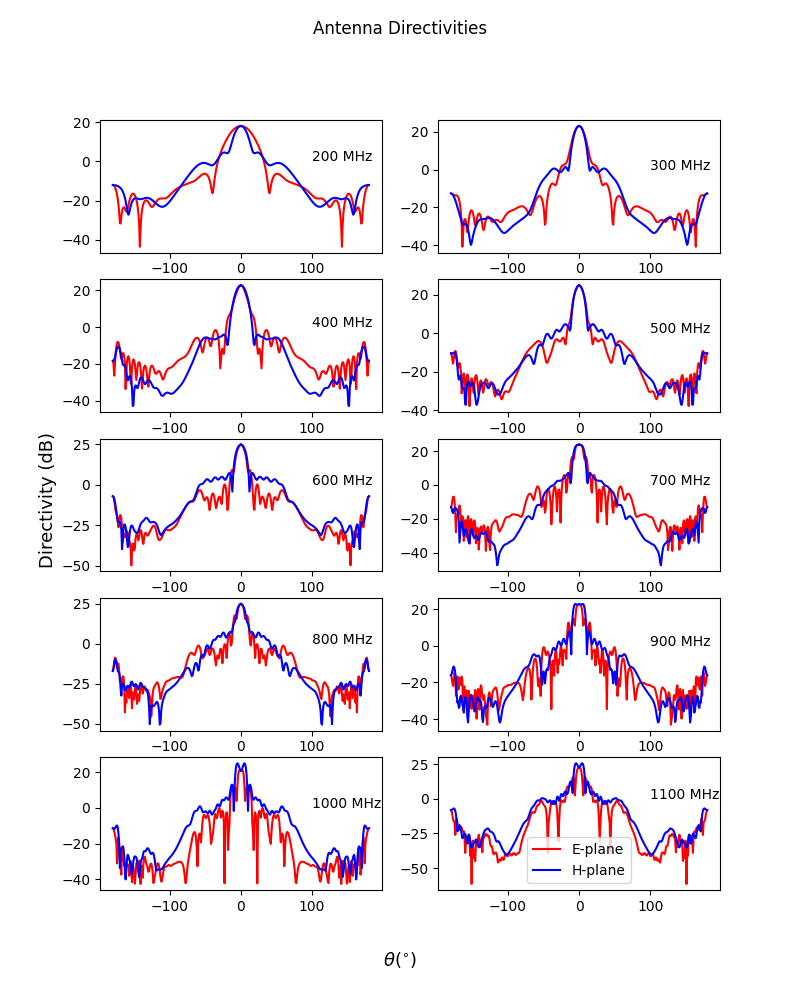}
    \caption{Directivities in the E and H-planes for the proposed antenna system}. These patterns were simulated using the $f/d = 0.216$ reflector with an elliptical collar. Note that the beam patterns of the minimum systematics feed look similar to these. However, the directivity at the horizon tends to be lower.
    \label{fig:beams}
\end{figure}
Some example beam patterns can be seen in Figure \ref{fig:beams}. These patterns were computed using an $f/d=0.216$ dish with an elliptical collar.  One will note that the value of the directivity at $90^{\circ}$ is about $40\,$dB to $50\,$dB below its peak value. This behavior is desirable as low directivity towards the horizon corresponds to lower mutual coupling between antenna elements. 

One will also notice however, that a there is a dip in gain at zenith for the $1000$\,MHz and $1100$\,MHz plots. This feature is not particularly concerning with respect to power spectrum measurements, as one is still sampling the same patch of the uv plane whether the null is present or not. So, as long as the beam is calibrated well, it should not be an issue. However, the complexity of these beams is concerning for calibration, since a complicated beam pattern will be more difficult to measure whether using drones or point source references.

It should be noted that this dip is due to phase center variation with respect to frequency. It can be fixed by adjusting the location of the feed with respect to the focus of the dish. However, such adjustments will come at the expense of phase efficiency at low frequencies.


\begin{wstable}[h]
    \centering
    \caption{Performance when the minimum systematics feed is paired with different reflector designs.}
    \begin{tabular}{@{}p{1.4cm} p{2.55cm} p{2.5cm} p{2.55cm} p{2.55cm} p{2.60cm}@{}} \toprule 
       & $\mathbf{S_{11}}$ \textbf{(dB)} & \textbf{Directivity (dBi)}& \textbf{Maximum Sidelobe, E-plane (dB)} & \textbf{Maximum Sidelobe, H-plane (dB)} & \textbf{$\mathbf{\eta_{sp}}$} \\ 
       \colrule\\
        
        \textbf{f/d=0.375} 
        & -7.6 at 200\,MHz 

        -14.5 at 600\,MHz 

        -35.8 at 1100\,MHz 

        & 20.2 at 200\,MHz 

        28.3 at 600\,MHz 

        31.1 at 1100\,MHz 

        & -24.4 at 200\,MHz 

        -15.8 at 600\,MHz 

        -20.1 at 1100\,MHz 

        & -20.0 at 200\,MHz 

        -15.6 at 600\,MHz 

        -18.7 at 1100\,MHz 

        & 0.929 at 200\,MHz 

        0.932 at 600\,MHz 

        0.927 at 1100\,MHz 
        \\ \colrule

        \textbf{f/d=0.216 without collar} & -7.0 at
        200\,MHz
        
        -15.5 at 600\,MHz
        
        -18.1 at 1100\,MHz
        
        & 18.7 at 200\,MHz 

        24.8 at 600\,MHz 

        24.0 at 1100\,MHz 

        & -16.9 at 200\,MHz 

        -16.2 at 600\,MHz 

        -9.4 at 1100\,MHz 

        & -13.9 at 200\,MHz 

        -14.7 at 600\,MHz 

        -14.3 at 1100\,MHz 

        & 

        0.985 at 200\,MHz 

        0.996 at 600\,MHz 

        0.998 at 1100\,MHz 

        \\ \colrule

        \textbf{f/d=0.216 with collar} & 
        -7.0 at 200\,MHz 
        
        -15.0 at 600\,MHz 
        
        -18.3 at 1100\,MHz

        & 19.4 at 200\,MHz 

        24.2 at 600\,MHz 

        24.0 at 1100\,MHz 

        & -19.3 at 200\,MHz 

        -15.8 at 600\,MHz 

        -10.1 at 1100\,MHz 

        & -14.5 at 200\,MHz 

        -15.7 at 600\,MHz 

        -15.0 at 1100\,MHz 

        & 0.989 at 200\,MHz 

        0.998 at 600\,MHz

        0.998 at 1100\,MHz 
        
         \\ \botrule
    \end{tabular}
 \label{tab:reflectors}
\end{wstable}

\subsection{Feed}
The feed designs presented in this paper are of a Vivaldi style. Such a design was chosen for its wide bandwidth, high directivity, and `openness'. It is easier to achieve high spectral smoothness with an open design since there are no enclosed spaces in which resonances might form. Moreover, the Vivaldi geometry minimizes aperture blockage, which is especially important in this context where the dish size is only $D \approx 4\lambda$ at the longest wavelength. 

\begin{figure}
    \centering
    \includegraphics[scale=0.23]{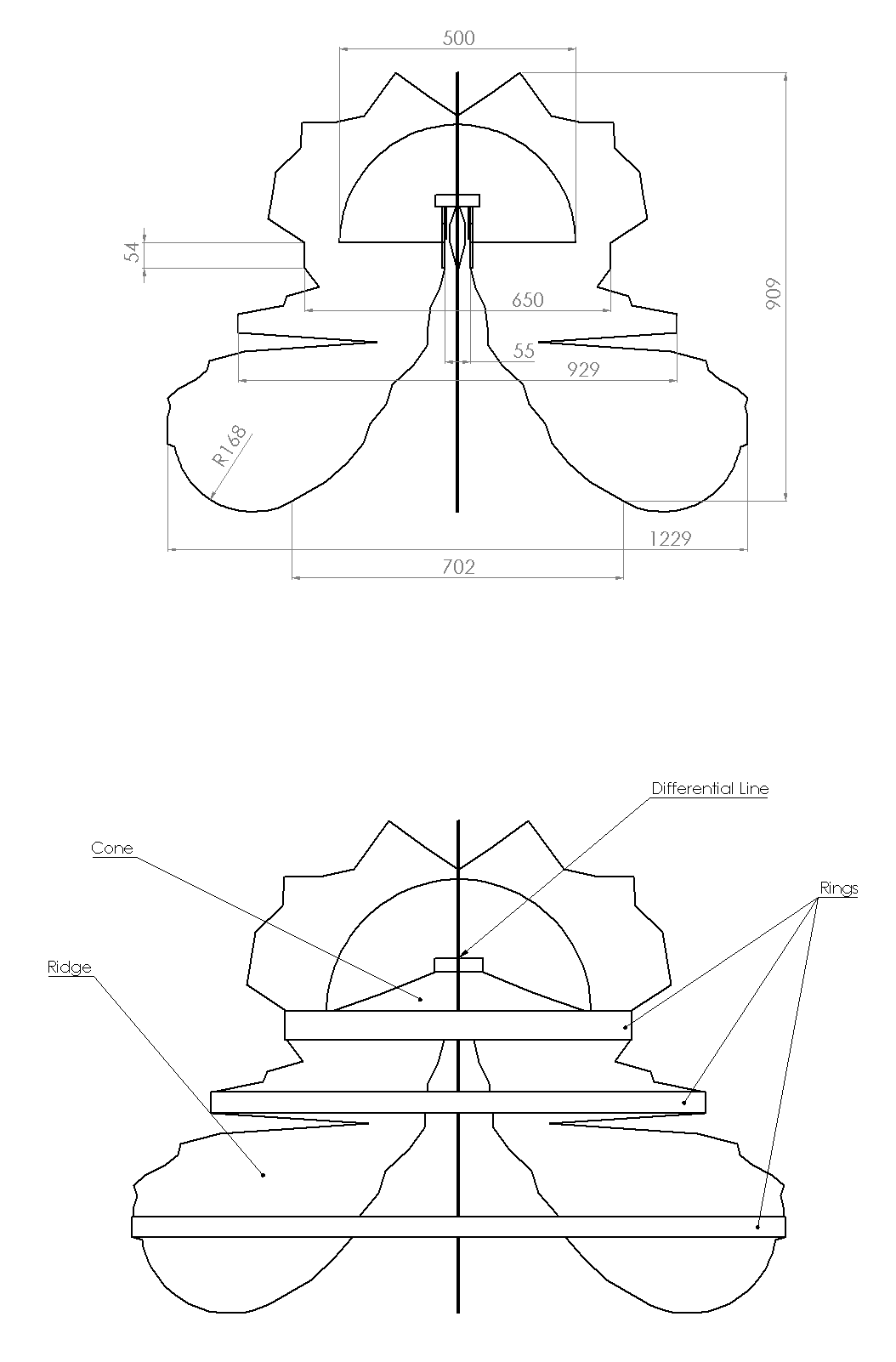}
    \caption{Top shows the minimum systematics feed with the cone and rings removed in order to make certain dimensions easier to see. All dimensions are in \,mm. Note that the ridges of the feed are $3.3$\,mm thick. The bottom image shows the minimum systematics feed with certain important pieces labeled. The proposed feed is identical to the minimum systematics feed, but with the rings removed.}
    \label{fig:feed}
\end{figure}

Figure \ref{fig:feed} presents a drawing of the minimum systematics feed. The proposed feed is identical to the minimum systematics feed, except it does not include the rings. All feed designs considered in this paper are shown in Figure \ref{fig:feed_versions}. In particular, the proposed feed is shown in the bottom left panel of the figure. Although not presented, horn antennas were also simulated but provided poor performance due to aperture blockage.

\subsubsection{Ridges}
\begin{figure}
    \centering
    \includegraphics[scale=0.55]{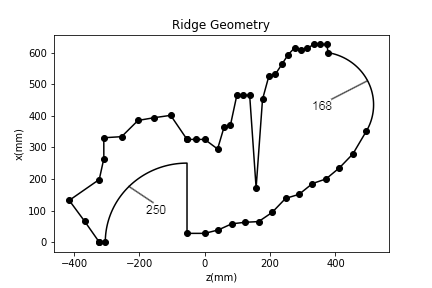}
    \caption{A drawing of one of the four ridges used on the proposed feed. $r = 168\,$mm refers to the radius of the circular flare used at the aperture. Points denote the locations $(z_{i},x_{i})$ used in the optimization.}
    \label{fig:drawing}
\end{figure}

This section describes the `ridges' of the feed, which are shown in Figure \ref{fig:drawing}.
As one can tell from the figure, the proposed feed was designed with a numerical optimizer.

In particular, the CMA-ES optimization algorithm was used \cite{hansen2003ecj}. This algorithm is built-in to CST.  The design of these ridges began as a more traditional Vivaldi shape, below called the `vanilla design', with an exponential flare inside and flat section outside. The profile was then broken into line segments connecting points $(x_{1},z_{1}),(x_{2},z_{2}),...$. The CMA-ES optimizer was then used to find the optimal values for the points $(z_{i},x_{i})$. Note that only $x_{i}$ was optimized for certain points and only $z_{i}$ for others.

Allowing the ridge geometry to vary in this way was found to minimize sidelobes and backlobes, increasing spill efficiency ($\eta_{s}$). Changing the ridge geometry from the vanilla to jagged design was found to degrade delay spectrum performance to some extent. However, it was found that this difference is negligible compared to the non-smooth contribution from dish-feed reflections. Moreover, the addition of the absorber-lined rings described in Section \ref{subsection:rings} helps somewhat with spectral-smoothness. So, the final feed design performs comparably to the vanilla design (shown in the bottom right of Figure \ref{fig:feed_versions}) as far as spectral smoothness is concerned.   

\subsubsection{Differential Line}
\begin{figure}
    \centering
    \includegraphics[scale=0.25]{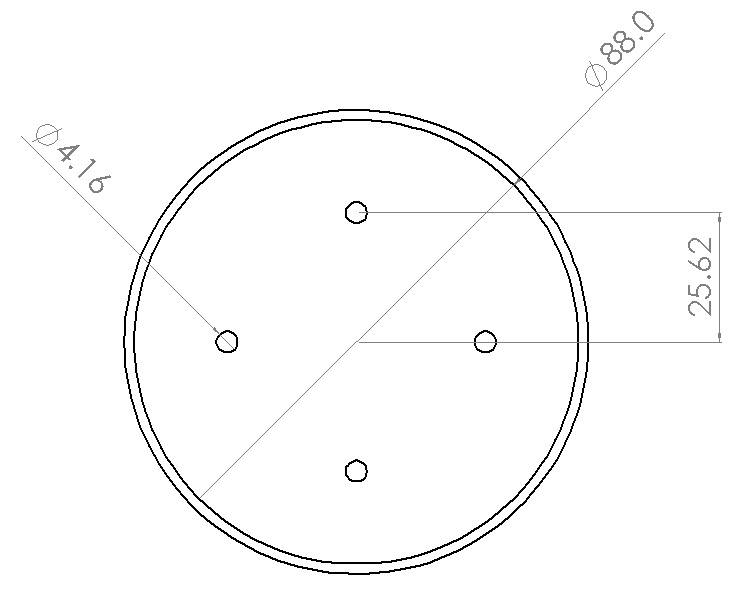}
    \caption{The differential line used on the feed. The line is $25.7$\,mm long.}
    \label{fig:line}
\end{figure}

 The output signals are carried from the antenna to amplifiers via a differential line, which can be seen in Figure \ref{fig:line}. Differential lines have been used on designs such as the horn described in \citet{beukman}, and the HERA Vivaldi described in \citet{hera_viv}.  This choice eliminates the need for baluns, thus reducing losses and lowering $T_{sys}$. It has also been shown to improve beam patterns \cite{beukman}. Moreover, it introduces the possibility of using common modes for beam pattern shaping \cite{beukman}.

This differential line is filled with a PTFE ``puck" of diameter $88\,$mm in order to decrease the impedance. The pins have diameter $4.2\,$mm. The line was chosen to be $26\,$mm long. These values were determined by the optimizer. The resulting line has impedance $Z_{0} \approx 200\,\Omega$.

Such an impedance is quite high and not ideal for impedance matching to the amplifier. The reason the impedance is high comes from the wide bandwidth of the feed. A lower impedance requires a larger diameter differential line. Such a line would allow $TE$ and $TM$ modes to appear at the higher frequencies in the band. It should be noted, though, that one could lower this impedance by replacing the PTFE with a dielectric with an even higher index. For instance, using Taconic RF-35, which has $\epsilon = 3.5$, would reduce the impedance to $\approx150\,\Omega$. However, it would be best to pair this antenna with an impedance matching circuit between the feed and amplifier.  This device would be similar to the HERA's front end module (FEM) \cite{hera_viv}.

\subsubsection{Tabs}

\begin{figure}
    \centering
    \includegraphics[scale=0.25]{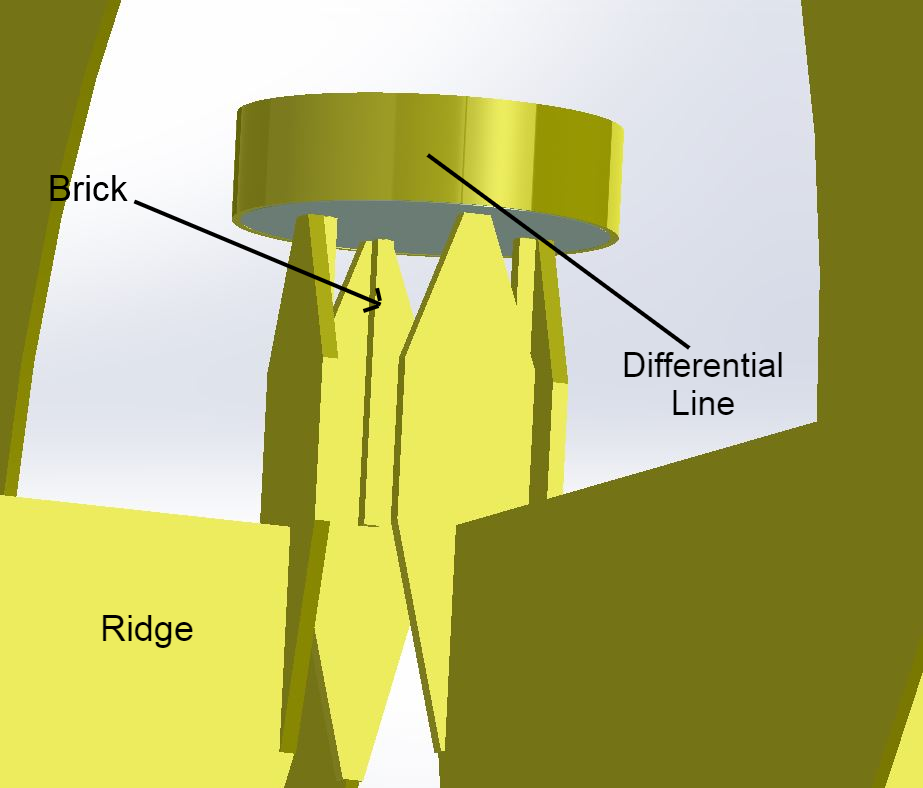}
    \caption{An illustration of the transition from the differential line to the ridges.}
    \label{fig:tab_trans}
\end{figure}
\begin{figure}
    \centering
    \includegraphics[scale=0.3]{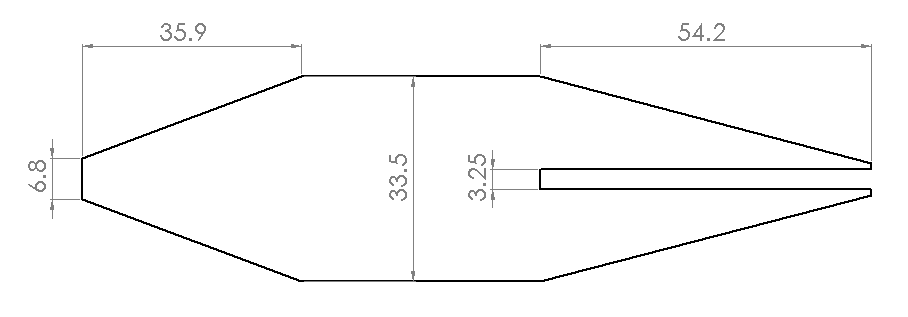}
    \caption{A drawing of one of the tabs used to transition from the differential line to the ridges, as seen in Figure \ref{fig:tab_trans}. Dimensions are in \,mm and the tabs are $4$\,mm thick.} 
    \label{fig:tabs}
\end{figure}
\begin{figure}
    \centering
    \includegraphics[scale=0.3]{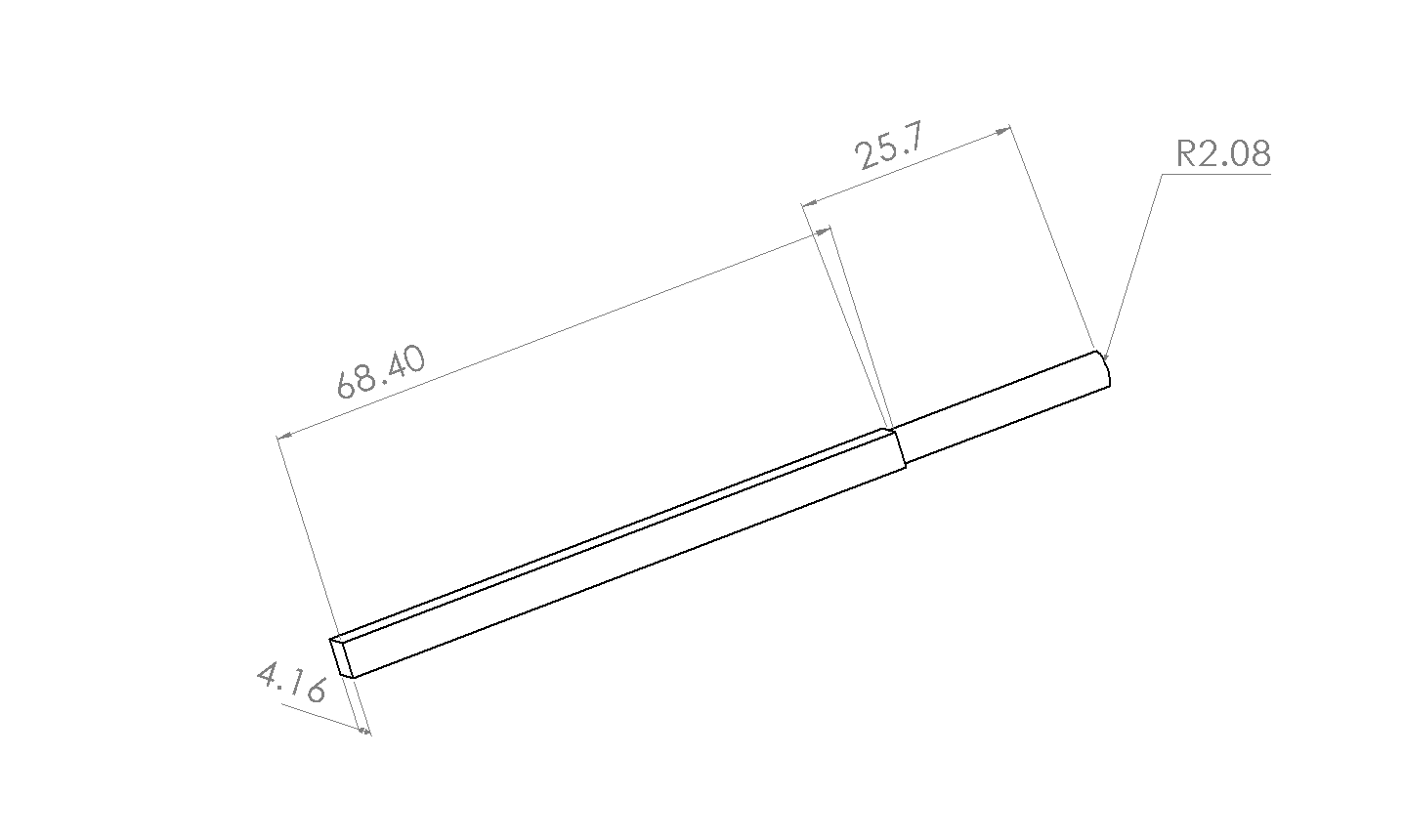}
    \caption{A drawing of the brick and pin used in the transition section and differential line respectively. One can see these pieces in the context of the feed in Figure \ref{fig:tab_trans}. The brick fits into the slot of the tab, while the pin is used as part of the differential line.}
    \label{fig:pin_and_brick}
\end{figure}
During optimization, it was found that the ideal thickness for the ridges on the feed was around $33\,$mm. In order to reduce weight and cost, the thickness of the ridges was reduced to about $3.3$\,mm. Reducing the thickness of the ridges increased the impedance of the feed, leading to a poor impedance match. In order to fix the issue, tabs were added, giving a smooth impedance transition from the differential line to the throat of the feed. A similar approach was used in \citet{hera_viv}. An illustration of this transition can be seen in Figure \ref{fig:tab_trans}. In this transition, the pins of the differential line are connected to rectangular ``bricks". These bricks are $4.2$\,mm $\times$ $4.2$\,mm $\times$ $68$\,mm and are attached to the tabs. The tabs then connect to the ridges, providing a smooth impedance transition. Another view of the tabs, with dimensions, can be seen in Figure \ref{fig:tabs}. By using the tabs, a good impedance match was achieved with only slight degradation in spillover efficiency compared to when thicker ridges were used. 

\subsubsection{Cone}
\begin{figure}
    \centering
    \includegraphics[scale=0.25]{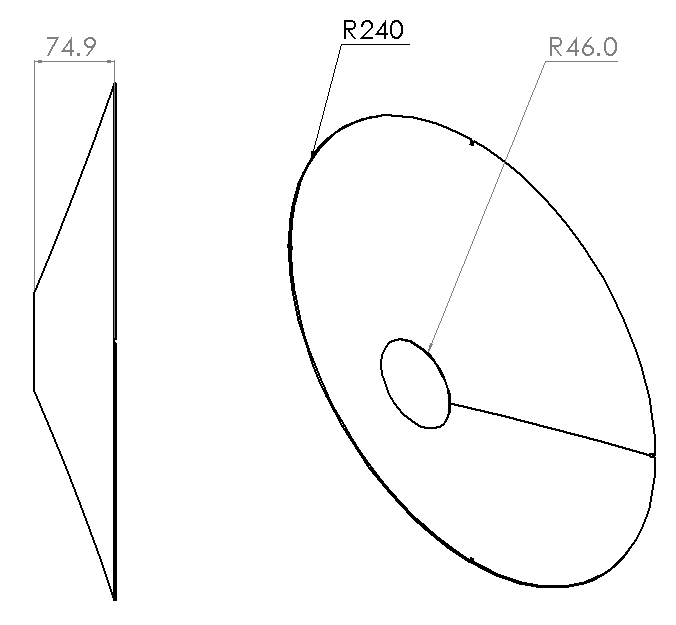}
    \caption{A drawing of the feed's cone, with dimensions. Our simulations used $1.7$\,mm as the thickness for the cone, but this dimension is not important for performance.}
    \label{fig:cone}
\end{figure}
To further improve beam patterns, a cone was introduced behind the throat of the ridges. This cone can be seen in Figures \ref{fig:feed} and \ref{fig:cone}. The cone is meant to prevent energy from radiating away from the differential line before reaching the ridges. The taper of the cone is intended to give a smooth impedance transition from the line to the throat section of the ridges. 
As shown below in Section \ref{subsection:comparison}, the cone degrades the $S_{11}$ at low frequencies by increasing the impedance. However, it also increases spill efficiency ($\eta_{s}$), which is a more important feature, since it corresponds to lower mutual coupling levels. Moreover, the presence of the cone was found to degrade spectral smoothness to some extent. However, as with the jagged ridge geometry, this is no longer the case once the rings are added to the design.   
\subsubsection{Rings} \label{subsection:rings}
Lastly, rings were introduced to the design to improve its beam patterns. The incorporation of the rings came from the desire to decrease cross-coupling. In order to decrease cross-coupling, one would like minimize the beam pattern at angles toward the horizon. This sort of behavior is achieved by lowering illumination at the edge of the dish, i.e. making the feed's beam pattern narrower. In order to make the beam narrower,  it was necessary to increase the size of the feed or add additional structure. It was found that the increased aperture blockage from the larger feed degraded performance. Wrapping a metal sheet around the feed to turn it into a quad-ridge horn was found to greatly improve the feed's beam patterns. However, as expected, this aperture blockage degraded performance. The compromise reached was to introduce the rings seen in Figure \ref{fig:feed}. These rings reduce the edge taper on the reflector while introducing minimal additional aperture blockage. 

The geometry of the feed had already been optimized before the rings were introduced, so rings were added to locations on the ridges that were amenable to them. The ideal number and placement of the rings was found through trial and error.

\begin{figure}[h]
    \centering
    \includegraphics[scale=0.7]{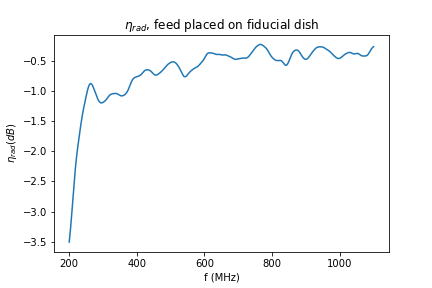}
    \caption{Radiation efficiency when the feed with rings is placed over a dish with $f/d = 0.216$ and an elliptical collar.}
    \label{fig:rad_eff}
\end{figure}

Note, however, that an absorber was added to the rings in order to prevent a resonance from occurring. For the full-sized feed, our proposed design uses a flexible ferrite absorber. In particular, model M6 from Fair-Rite\footnote{\url{http://www.fair-rite.com/flexible-ferrite/}} was used in the simulations. The simulations described were performed using a layer of absorber $1.5$\,{mm} thick on the largest ring and $0.5$\,{mm} thick on the other two rings. The absorber does degrade radiation efficiency, as can be seen in Figure \ref{fig:rad_eff}.
Although not shown, simulated radiation efficiency was nearly perfect in the absence of the absorber. Note that the radiation efficiency is poorest at the lowest end of the band, but increases to $-1$\,{dB} by $250$\,{MHz}.

\begin{figure}[h]
    \centering
    \includegraphics{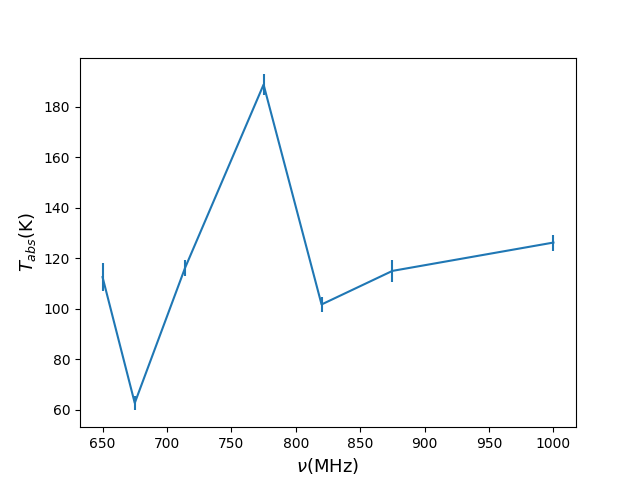}
    \caption{Noise contribution of the absorber in the  fabricated minimum systematics feed.}
    \label{fig:noise_temp}
\end{figure}

It should also be noted here that the presence of the absorber increases the noise temperature of the antenna system. The noise contribution was measured for the fabricated 4/13 scale minimum systematics feed. In particular, the noise of the feed was measured at room temperature and submerged in liquid nitrogen. The noise temperature was then obtained by taking the difference of the two measurements while accounting for amplifier gain and impedance matching. The results of these measurements are presented in Figure \ref{fig:noise_temp}. Note that only frequencies up to 1\,GHz are presented due to the limitations of the low noise amplifiers available to us. Also, the lowest frequency measured was 650\,MHz, as this corresponds to the lowest frequency in the band of the 4/13 scale feed. From these measurements, it was found that the typical noise contribution of the absorbers was $\approx 120$\,K. Although these measurements were taken for the scaled down feed, it is safe to assume results would be similar for the full scale feed, as simulations showed similar beam patterns and S-parameters for the two versions.


\subsection{Optimization of the Feed} \label{subsection:opt}
Since this feed was optimized with PUMA in mind, the dimensions shown in the following sections are set to optimize performance from $200\,{\rm MHz}$ up to $1100\,{\rm MHz}$. 

The optimizer's error functions were defined as 
\begin{equation}
    \mathcal{E} = a\mathcal{E}_{s}+b\mathcal{E}_{S11}+c\mathcal{E}_{d}.
\end{equation}
In this equation, $\mathcal{E}_{s}$ considers spill efficiency, $\mathcal{E}_{S11}$ considers impedance matching, and $\mathcal{E}_{d}$ is used to keep track of delay spectrum performance. The weights $a,b,c$ were typically chosen to be $\approx 1$, but their values were sometimes adjusted if some aspect of performance was lagging behind the others. 

The error for the $S_{11}$ term is defined via
\begin{equation}
    \mathcal{E}_{S11} = \frac{1}{N}\Sigma_{n=1}^{N} \Theta(S_{11}[n] +10) [S_{11}[n]+10]^{2}. 
\end{equation}
In this formula, $n$ indexes the frequency samples in the simulation, $S_{11}$ is given in dB and $\Theta$ is the Heaviside step function. In other words, the goal was to keep $S_{11}$ below $-10\,$dB.

The error in the spill term $E_{S}$ is defined as 
\begin{equation}
    \mathcal{E}_{s} = \Sigma_{n} \Theta(0.93 - \eta_{s}[n])(0.93 - \eta_{s}[n]).
\end{equation}
In this formula, $\eta_{s}$ is defined as the portion of the feed's beam illuminating the dish. (Note that this is not the same as $\eta_{sp}$, which is defined as the portion of the dish's beam illuminating the sky.) The spill $\eta_{s}$ is calculated assuming a dish with an $f/d$ ratio of 0.25. Also, note that $n$ is an index used to label the frequencies at which $\eta_{s}$ was estimated. During optimization, it was found that the fine-tuned ridge geometry was not capable of improving $\eta_{s}$ over the entire $200$ to $1100\,{\rm MHz}$ bandwidth. As such, $\eta_{s}$ was only considered for frequencies $200\,{\rm MHz} < f < 475\,{\rm MHz}$. This choice was made to maximize performance at the frequencies most important for PUMA's science goals.
Finally, the delay error was defined as
\begin{equation}
    \mathcal{E}_{d} = \Theta(\Delta t - 17)(\Delta t - 17),
\end{equation}
where $\Delta t$ is the time in nanoseconds that it takes for the energy in the system to drop by $30$\,dB when using the CST time domain solver. This ensures that no resonances are present in the feed. The $17$\,ns value was chosen based on a candidate design that was considered to have adequate delay spectrum performance.

\subsection{A Performance Comparison} \label{subsection:comparison}
\begin{figure}
    \centering
    \includegraphics[scale=0.1]{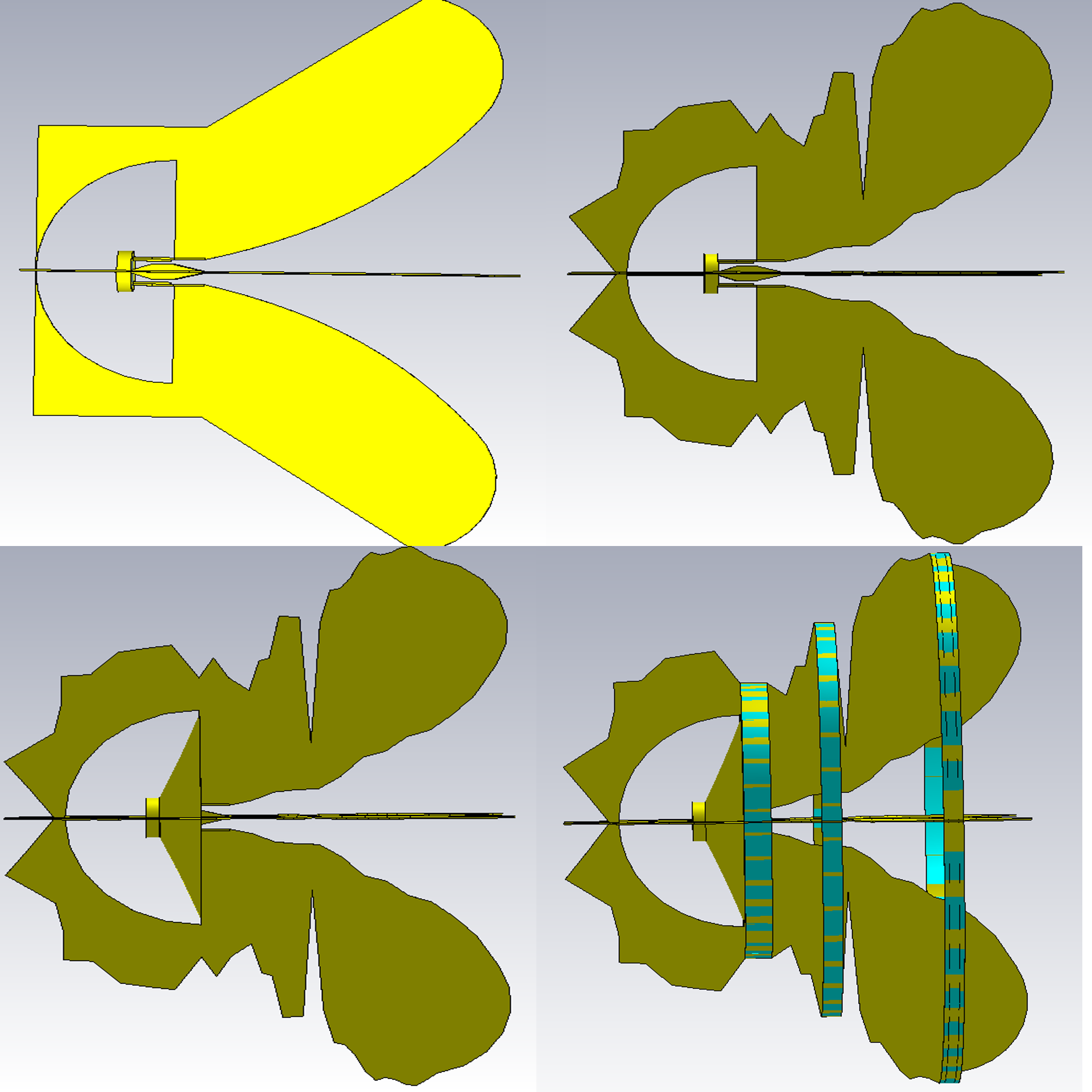}
    \caption{Pictures of feed versions. In reading order: vanilla feed, jagged ridges no cone, proposed feed and minimum systematics feed.}
    \label{fig:feed_versions}
\end{figure}

\begin{figure}[h]
    \centering
    \includegraphics[scale=0.57]{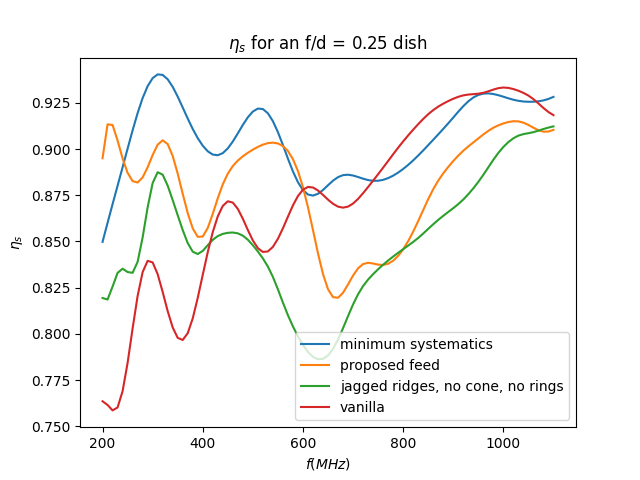}
    \caption{Comparisons of simulations of $\eta_{s}$ as a function of frequency for different versions of the feed.}
    \label{fig:spillover}
\end{figure}

\begin{figure}[h]
    \centering
    \includegraphics[scale=0.57]{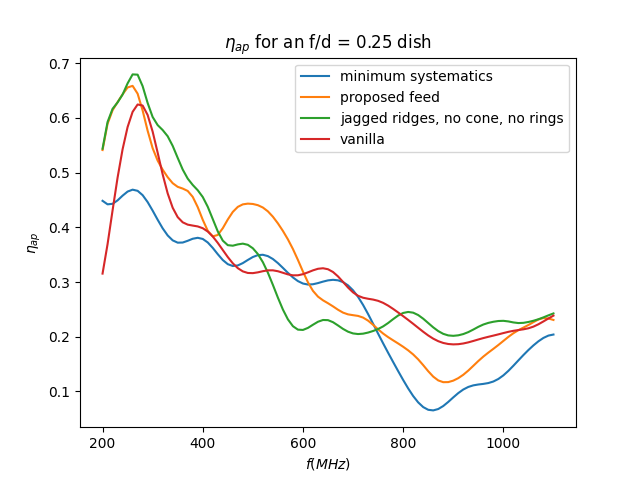}
    \caption{Comparison of simulated aperture efficiencies on a $f/d=0.25$ dish as a function of frequency.}
    \label{fig:apeff}
\end{figure}

\begin{figure}[h]
    \centering
    \includegraphics[scale=0.57]{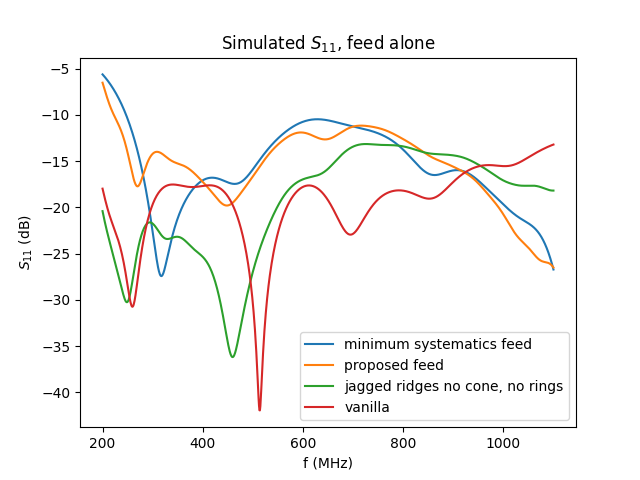}
    \caption{Simulated $S_{11}$ values for different versions of the feed. The values are simulated without the dish, and assuming a reference impedance of $200\,\Omega$.}
    \label{fig:sparam_feed}
\end{figure}

\begin{figure}[h]
    \centering
    \includegraphics[scale=0.4]{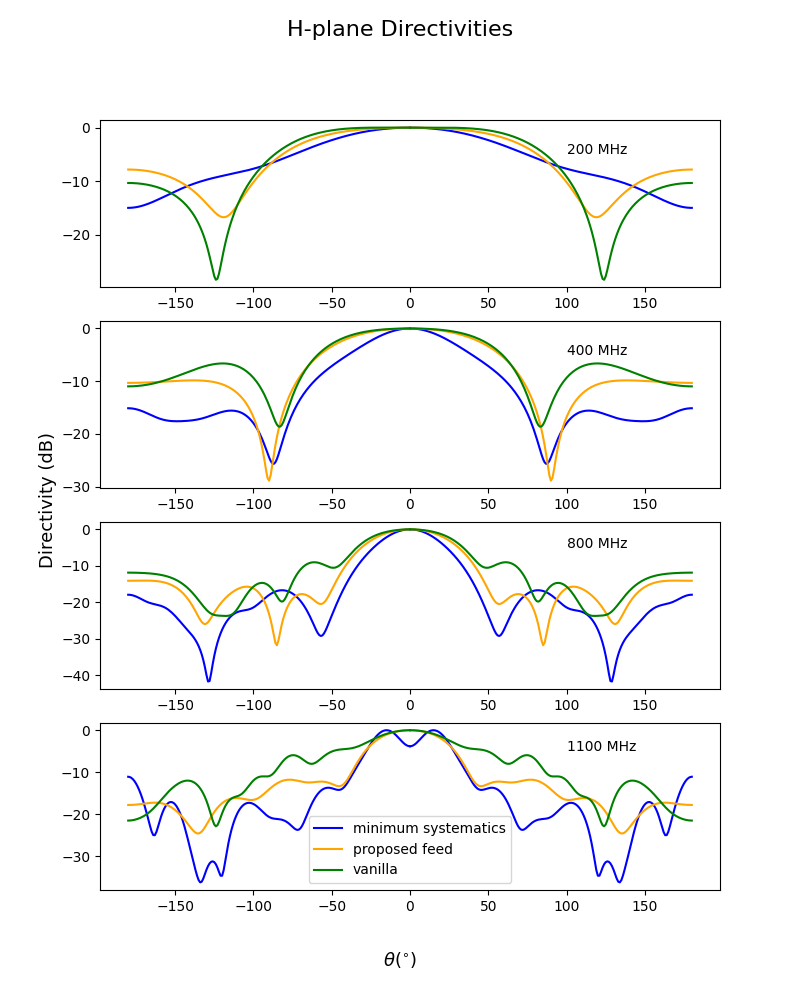}
    \caption{Simulated peak-normalized H-plane directivities for different versions of the feed.} 
    \label{fig:feed_beams_h}
\end{figure}

In this subsection, plots comparing the simulated performance for different feed designs are shown to illustrate the benefits of the design features added. These feed versions can be seen in Figure \ref{fig:feed_versions}. 
In particular, $\eta_{s}$, which is the fraction of the feed's beam that illuminates the dish, $\eta_{ap}$, the $S_{11}$ (of the feed alone), and some beam patterns are shown in Figures \ref{fig:spillover}, \ref{fig:apeff},\ref{fig:sparam_feed}, and \ref{fig:feed_beams_h}, respectively.

In Figure \ref{fig:spillover}, $\eta_{s}$ is plotted for various iterations of the feed. ``vanilla" (Figure \ref{fig:feed_versions}) refers to a version of the feed with no cone and a simple exponential taper and straight lines for the outside and back part of the profile. One can see that the introduction of the jagged profile improves spillover efficiency at low frequencies but degrades it above $\approx 400$\,MHz. As mentioned in Section \ref{subsection:opt}, this is because jagged ridges can't improve the beam patterns for all frequencies. So, the jagged ridges were optimized to improved performance in the lower part of the band where the most cosmological information is contained. 

Introducing the cone (the proposed feed) reduces spillover at all frequencies, and is especially effective for $f<300$\,MHz. Introducing the rings (minimum systematics) improves  the spillover efficiency even further, and fixes the problems caused by the jagged ridges at higher frequencies. 

Figure \ref{fig:apeff} shows simulated aperture efficiencies for the various versions of the feed design. This plot shows that the introduction of the curved ridges greatly improves the aperture efficiency at the lower end of the band. On the other hand, the introduction of the rings seems to degrade the efficiency of the minimum systematics feed. One will also notice that the aperture efficiency decreases significantly at frequency increases. This is due to the fact that the feed geometry is large compared to such short wavelengths, meaning that the dish is located in the nearfield of the feed. Although not shown, predictions of aperture efficiency based on feed farfield patterns agree with Figure \ref{fig:apeff} at low frequencies but begin to disagree significantly as frequency increases. Despite the low aperture efficiency, these feed designs can still achieve adequate sensitivity at high frequencies due to their high spillover efficiencies.

Figure \ref{fig:sparam_feed} shows simulated $S_{11}$ values for different versions of the feed. This plot shows that versions of the feed without the cone have $S_{11} \lessapprox -15$\,dB for the entire band. The introduction of the cone raises the $S_{11}$ above $-10$\,dB for the frequencies below $f \approx 250$\,MHz. This is due to the fact that the cone raises the impedance of the antenna at these frequencies. This slightly higher $S_{11}$ should not cause significant problems, however, since receiver noise is subdominant to the sky temperature at these frequencies \cite{cosmic_visions}. Aside from that, the $S_{11}$ stays below the goal of $-10$\,dB for the rest of the band. Lastly, note that introducing the rings only makes a slight difference to the $S_{11}$. 

\begin{figure}[h]
    \centering
    \includegraphics[scale=0.55]{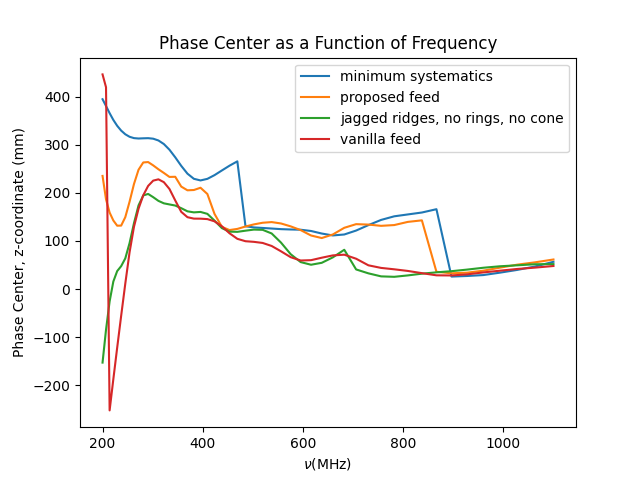}
    \caption{A plot showing the boresight coordinate of the phase center as a function of frequency.}
    \label{fig:phase_center}
\end{figure}
Figure \ref{fig:phase_center} shows how the phase center varies with frequency for different feed designs. These phase centers were computed by CST. The ``boresight" setting in CST was used, with a $90\,^{\circ}$ degree angle around the z-axis. The values shown here are the average of the phase centers computed in the E-plane and H-plane.

First,  note that the phase center \cite{beukman} of the minimum systematics design varies by about $400$\,mm across the band. Such variation means that achieving good phase efficiency for the minimum systematics feed may not be possible over the whole band. Also, note that the cone helps keep the phase center stable at low frequencies. This is due to the fact that the cone prevents a resonance from occurring in the backshort, which has a perimeter close to $\lambda/4$ at the longest wavelengths of operation. Moreover, the feed with rings seems to have three different regimes of behavior. At low frequencies, the phase center is located close to the largest ring. In the middle of the band, the phase center is closer to the middle ring. At low frequencies the phase center is close to the smallest ring. 

This phase center variation is responsible for the dip in gain at zenith seen in figure \ref{fig:beams} at high frequencies. For these beam patterns, the feed was placed $1500$\,mm above the dish, meaning that the phase center is located $200$\,mm from the focus. At low frequencies, this means that the phase center will be quite close to the focus. However, at higher frequencies the phase center will be about $150\,$mm from the focus, leading to higher phase errors. These errors are responsible for the dip in gain seen at zenith in Figure \ref{fig:beams}. In this paper, we have chosen the placement in order to optimize for performance at lower frequencies.


\begin{figure}[h]
    \centering
    \includegraphics[scale=0.4]{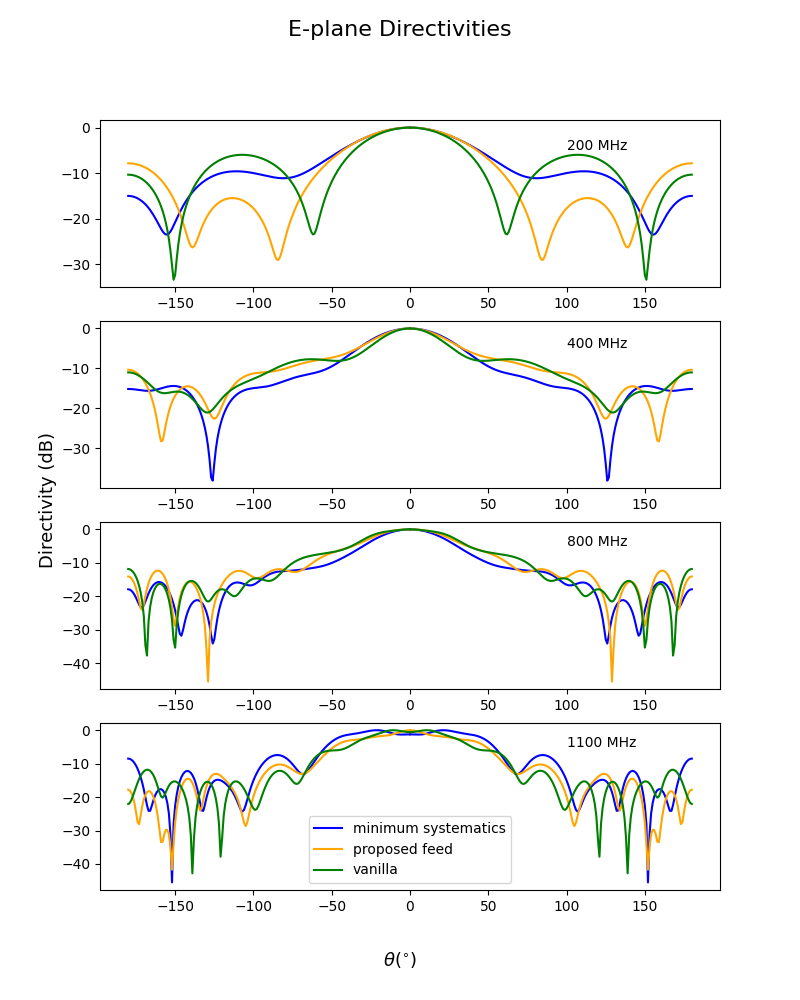}
    \caption{Simulated peak normalized E-plane directivities for different versions of the feed.}
    \label{fig:feed_beams_e}
\end{figure}
Finally, consider the beam patterns of the feeds. H-plane patterns for the proposed feed, the jagged feed with a cone, and the vanilla feed can be seen in Figure \ref{fig:feed_beams_h}. Note first that the presence of the rings causes the beam pattern to taper off more quickly and smoothly. This feature is important since it will help suppress mutual coupling in the H-plane, where it tends to be most severe.  Moreover, note that the proposed feed has suppressed the backlobes. The smaller backlobes are a result of the rings used in that design. E-plane beam patterns can be seen in Figure \ref{fig:feed_beams_e}. For the E-plane, note that the red proposed feed performs the best at the very lowest end of the band. Moreover, it can be seen that the different designs perform similarly as frequency increases.

\section{Simulated Performance and Comparison to Requirements}
\label{section: sim perf}
In this section, relevant aspects of the antennas' performance are presented and compared to the requirements described in Section \ref{section:Design Requirements}. Note that these metrics were all evaluated with the feed placed on a $6$\,m diameter dish with $f/d = 0.216$. Since no such dishes were available for testing, estimated results based on simulations will be shown instead. Measured beam patterns and S-parameters for the proposed feed will be shown in Section \ref{section: fab}. 

\subsection{Delay Spectrum Performance}
Recall from Section \ref{subsection:smooth} that the delay response of the instrument can be characterized using $\tilde{R}(0,0,\tau)$, the Fourier transform of the power kernel at zenith. In order to compute this, the antenna is excited with a plane wave propagating along the boresight of the antenna. This simulation was performed using the time domain solver in CST, with a planewave excitation coming from above, the differential line of the feed terminated by a waveguide port, and a voltage monitor placed between two pins on the differential line. The time domain solver simulated up until $400\,$ns after the initial excitation. This $400\,$ns time was chosen in order to capture delay spectrum performance at large delays. The voltage present at the antenna's terminals is recorded in order to obtain values for ${R}(0,0,\nu)$. Then, two Blackman-Harris filters are applied in order to smoothly bring the power kernel to 0 at the edges of the band. These filters will be denoted as $W(\nu)$. This leads to a delay space power kernel 
\begin{equation}
    \tilde{R}(0,0,\tau) = \int d\nu R(0,0,\nu) W(\nu)^{2} e^{2\pi i \nu \tau}.
\end{equation}

\begin{figure}
    \centering
    \includegraphics[scale=0.4]{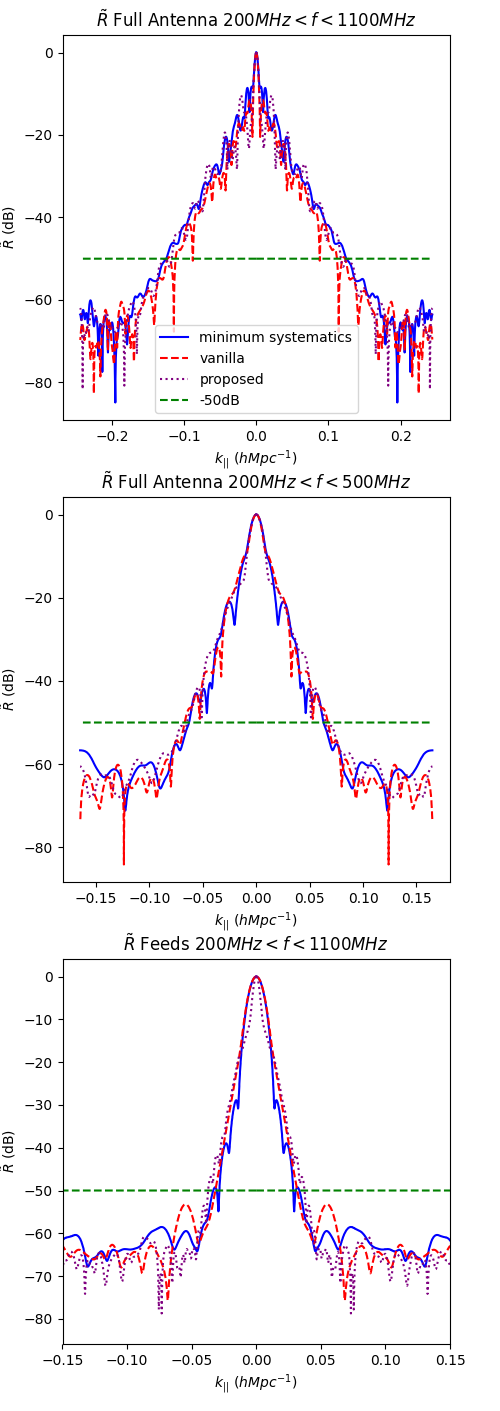}
    \caption{Delay spectrum: comparison of the power kernel of the full antenna for the minimum systematics, vanilla, and proposed feed. The top plot is for the full band. The middle plot was made using a bandpass that selects for frequencies $200\,{\rm MHz} < \nu < 500\,{\rm MHz}$. The bottom plot shows the power kernel of the feeds without the reflector present.}
    \label{fig:delay}
\end{figure}

Results for minimum systematics, proposed and vanilla feeds are shown in Figure \ref{fig:delay} in terms of $k_{||}$, which is the spatial wavenumber along the line of sight of the telescope. In cosmological contexts, one finds that $k_{||} \propto \tau$.

First, note from Figure \ref{fig:delay} that all three designs provide similar results in all three plots. So, it can be concluded that the presence of the cone, rings, and jagged ridges does not significantly worsen delay spectrum performance. Next, note that the top plot of the figure shows the power kernel dropping by $50\,$dB after a delay corresponding roughly to $k_{||} \approx 0.1\,{\rm hMpc^{-1}}$. Thus, roughly speaking, modes within $0.1\,{\rm hMpc^{-1}}$ of the wedge would be avoided in order to be sure that the foregrounds are not overpowering the HI signal. Readers should also note that lumps appear in the delay spectrum as $k_{||}$ increases. These lumps are due to reflections bouncing between the feed and the dish. In the middle plot, one can see that the lower end of the band performs a bit better, dropping by $50\,$dB after about $0.065\,{\rm hMpc^{-1}}$. This result is of interest as the band corresponds to redshifts $2 < z < 6$, which are particularly important for the PUMA experiment. Lastly, the bottom plot shows the delay spectrum for the feeds only. Readers will note that the kernel drops off quite quickly in this case, reaching $-50\,$dB within $k_{||} \approx 0.03\,{\rm hMpc^{-1}}$. This is due to the absence of dish-feed reflections, which worsen performance in the middle and top plots. 
As mentioned in Section \ref{section:Design Requirements}, the power kernel should drop $50\,$dB as quickly as possible. However, the presence of dish-feed reflections slows down this process, giving a $-50\,$dB width that corresponds to $0.1\,{\rm hMpc^{-1}}$. The use of a higher $f/d$ or off-axis design would help reduce these reflections. However, as seen in Saliwanchik et al. (2021), the increased mutual coupling from such a design would be worse than the dish-feed reflections one incurs with a deep dish \cite{hirax}.

\subsection{Mutual Coupling Performance} \label{subsection:mutual}
\begin{figure}
    \centering
    \includegraphics[scale=0.6]{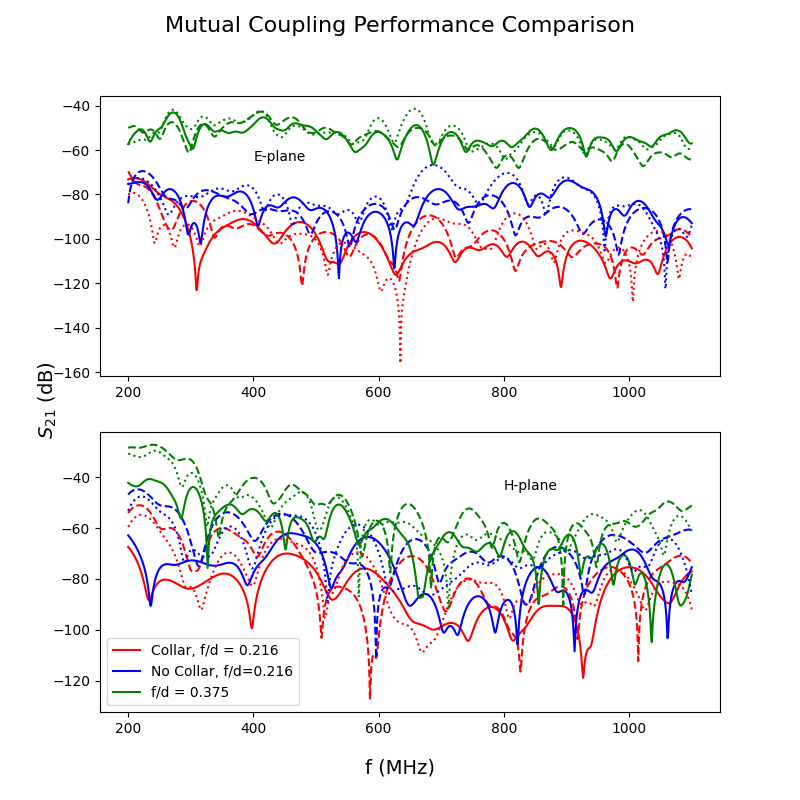}
    \caption{Mutual coupling: $S_{21}$ values for various antenna options. The top plot shows $S_{21}$ values calculated when each antenna is located in the E-plane of the other. The bottom plot shows $S_{21}$ values when the antennas are located in each other's H-plane. Colors label different dish choices. Dotted lines correspond to the proposed feed, solid lines correspond to the minimum systematics feed, while dashed lines are used for the vanilla feed.}
    \label{fig:coupling}
\end{figure}

In order to evaluate the proposed antenna's mutual coupling performance, the S-parameters for two antennas were considered. These antennas both point at zenith and are taken to be separated by $5.3 \lambda$ at the longest wavelength. Section \ref{subsection:XC} provides motivation on this setup. For these simulations, the vanilla, proposed and minimum sytematics feeds were considered. The results of these simulations are shown in Figure \ref{fig:coupling}. 

In the E-plane, the $f/d = 0.216$ no-collar dish has mutual coupling $20\,$dB to $40\,$dB lower than the shallower $f/d = 0.375$ dish. When the collar is used, the mutual coupling is $20\,$dB to $40\,$dB lower than the shallow dish when frequencies are between $200\,$MHz and $400\,$MHz. For the rest of the band, the $S_{21}$ is $40\,$dB to $60\,$dB lower when the deep with-collar dish is used rather than the shallow dish. 

In the H-plane, the use of a deeper dish also helps with mutual coupling. The deep, no-collar dish provides about a $10\,$dB to $20\,$dB improvement in the lower part of the band. Note that the presence of the collar doesn't help quite as much in the H-plane, providing only about a $5\,$dB improvement. 

Now, the various feed options will be compared. First, note that the proposed feed provides an advantage over the other two options in the E-plane at low frequencies for the deep dish with collar case. On average, we find the proposed feed providing about a $10\,$dB improvement over the vanilla feed for frequencies $\lesssim 300\,$MHz. Moreover, the proposed feed also provides some improvement over the vanilla feed in the H-plane at low frequencies, giving about a $10\,$dB improvement over the vanilla feed for frequencies $\lesssim 400\,$MHz. Notably, the minimum systematics feed does provide a coupling level about $20\,$dB better than the vanilla feed in the H-plane at the lower end of the band (from $200\,$MHz up to $400\,$MHz). Although these may sound like modest improvements at first, recall the lower end of the band is critical for PUMA's science goals. Thus, any improvement in that part of the band is especially important.

Recall also that the goal was to keep $S_{21}$ levels below $-100\,$dB. This goal is achieved in the E-plane for all designs for frequencies above approximately $600\,$MHz, when the deep dish with collar is used. Results do not quite meet the goal in the H-plane. The $S_{21}$ of the minimum systematics feed stays below about $-70\,$dB for the entire band, which, although an improvement, does not meet the goal. 

Recall that this simulation is meant to cover the worst case scenario, where the dishes are very close. For longer baselines, this coupling would certainly go down. However, shorter baselines cannot be ignored, since they provide information about large angular scales. 

Since the mutual coupling level is still not quite low enough at PUMA's redshifts of interest, it will be necessary to remove this effect during the data analysis stage. So far, techniques have been developed for the removal of mutually coupled signals from the data \cite{kerns} \cite{Parsons_2010}. These techniques rely on the fact that coupled receiver noise varies slowly with time compared to the sky signal. However, this type of approach doesn't work so well for NS-oriented baselines, for which the sky signal also varies slowly with time. Moreover, how to handle the effects of mutual coupling on antenna beam patterns is still an open question. Thus, new techniques are likely needed to control mutual coupling. 

\subsection{Sensitivity Performance}
\begin{figure*}
   \label{fig:sensother}
   \centering
   \includegraphics[scale=0.4]{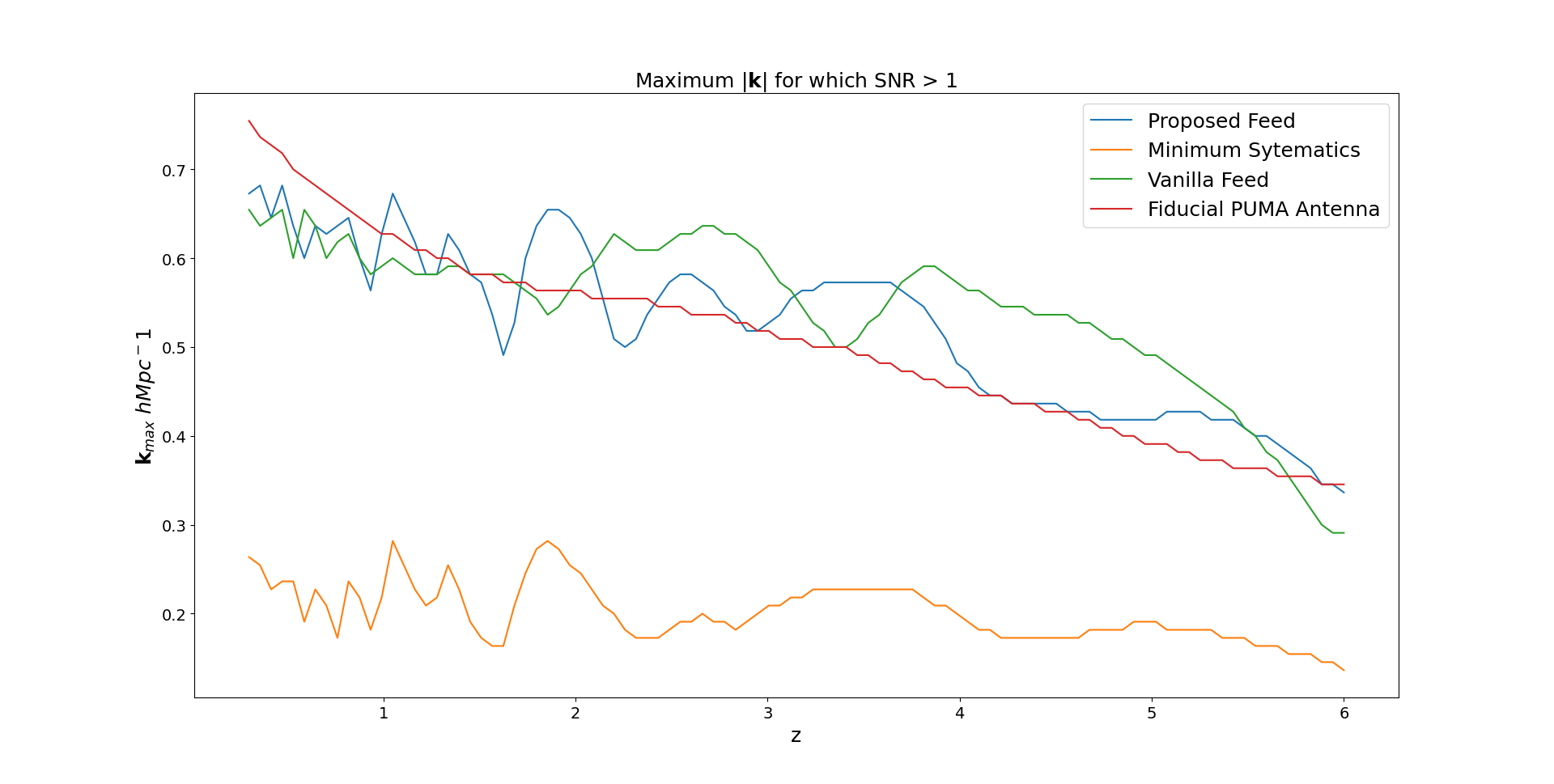}
   \caption{Maximum $|\mathbf{k}|$ for which ${\rm SNR}>1$. These calculations were performed assuming $\hat{k} \cdot \hat{n} = 0.5$.}
    \label{fig:sensitivity}
\end{figure*}

In order to evaluate the sensitivity of the antennas, the PUMA noise calculator \footnote{\url{https://puma.bnl.gov}} was used. This calculator takes in parameters describing the properties of a fiducial PUMA design. These parameters describe beam size, observing time, and fraction of the sky observed. It then produces an estimate for the power spectrum due to noise via Equation D4 given in \cite{cosmic_visions}. This equation scales as $\Omega_{p}$ as in Eq \ref{eqn:bull}. This contribution is computed for one $\mathbf{k}$ mode at a time. The calculator was edited, however, to use the formula for the noise power spectrum given in Eq \ref{eqn:parsons}. Sensitivity estimates are presented in Figure \ref{fig:sensitivity}. Note that all of these curves assume an $f/d=0.216$ dish with a collar included. Calculations were performed assuming a 32,000 element hexagonal array with $1.25$ years of integration time. The integration time was set assuming that PUMA would run for 5 years with 6 hours per day of observation. A value of 50\,K for was assumed for $T_{ampl}$. For the minimum systematics feed, we accounted for the noise contributed by the absorber by adding an additional term to Eq \ref{eq:tsys}. This term is given by 
\begin{equation}
    \frac{\eta_{s_{11}}T_{abs}}{\eta_{c}\eta_{sp}}.
\end{equation}
For these calculations, we assumed that $T_{abs}$, the noise temperature provided by the absorber, is $117$\,K. This value was obtained by averaging the data points we measured for Figure \ref{fig:noise_temp}. Note that this is a rough estimate, since we were unable to accurately measure the noise contribution at higher frequencies.

The beam patterns used for the various antenna designs come from CST simulations with a single antenna pointing at zenith. Note that sensitivity will decrease slightly for pointings away from zenith due to increased noise pickup from the ground. This plot shows the performance of the proposed feed when paired with different reflector designs. The performance of the idealized PUMA antenna described in \citet{cosmic_visions} is also shown. For the fiducial PUMA antenna, an Airy beam pattern and the parameters given in Section \ref{subsection:sensitivity} were used.    



Note in Figure \ref{fig:sensitivity} that the proposed feed and the vanilla feed do provide comparable sensitivity to the fiducial PUMA antenna, with somewhat worse performance at lower redshifts and somewhat improved performance at higher redshifts.


Lastly, one will also surely notice that the sensitivity provided by the minimum systematics feed is significantly lower than the other feed designs presented. This is due to the decrease in radiation efficiency and increase in noise caused by the absorber present on the rings. In the context of PUMA, this leads to significantly decreased sensitivity to smaller scale modes. However, this increase in noise temperature may not be so severe in the context of EOR telescopes, where the sky temperature can be up to $1000\,$K, which is significantly higher than the $\approx 100\,$K provided by the absorber.

\subsection{Cross-Polarization}
\label{subsection:crosspol}
\begin{figure}
    \centering
    \includegraphics[scale=0.75]{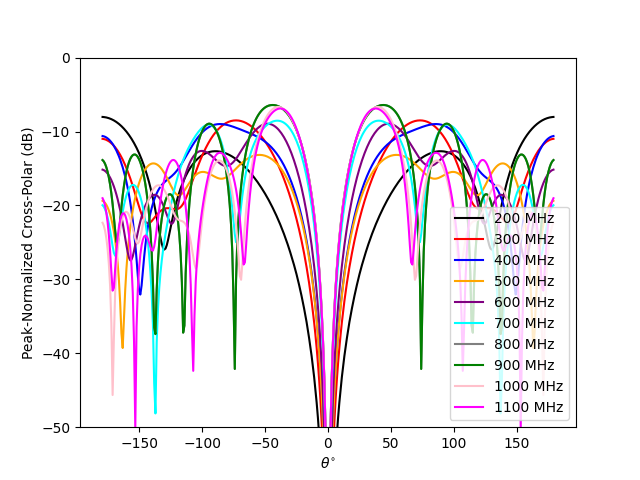}
    \caption{Peak-normalized cross-polar beam patterns of the proposed feed in the D-plane.}
    \label{fig:feed_cross}
\end{figure}

Although not considered for the performance requirements of the antenna, the cross-polarization properties will now be presented. Figure \ref{fig:feed_cross} shows the peak-normalized cross-polarized pattern of the proposed feed in the D-plane, where it is most severe. The peak-normalized patterns are given in dB, as 
\begin{equation}
    20\mathrm{log_{10}}\left(\frac{|E_{xp}(\theta,\phi)|}{\mathrm{max}(|E_{co}|)}\right),
\end{equation}
where the $E$ functions refer to the electric field patterns of the feed. For this paper, we use the Ludwig-3 convention for cross-polarization \cite{ludwig} \cite{beukman}. These values vary significantly with $\theta$, with the highest levels being about 10\,dB below the peak value. These peak values are comparable to those shown in \citet{beukman}.

\begin{figure}
    \centering
    \includegraphics[scale=0.75]{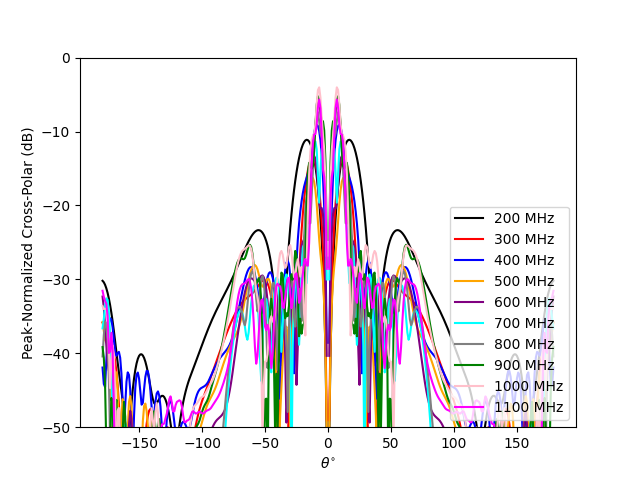}
    \caption{Peak-normalized cross-polar beam patterns of the proposed antenna system in the D-plane.}
    \label{fig:cross_dish}
\end{figure}

Figure \ref{fig:cross_dish} shows peak-normalized cross-polar beam patterns for the proposed antenna system, including both dish and feed. 
In this figure, one finds cross-polar beam patterns approaching $0$ at angles close to the boresite 
These patterns peak at angles close to the edge of the main-lobe, with the sharpest peaks located at the upper end of the band.

\begin{figure}
    \centering
    \includegraphics[scale=0.75]{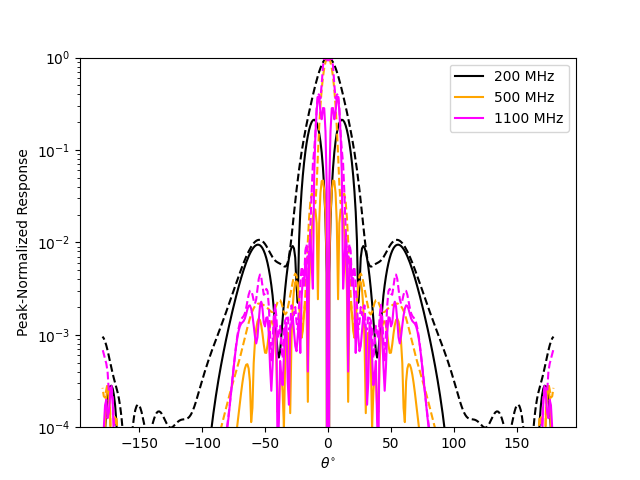}
    \caption{Peak-normalized $R$ values for three frequency channels. Solid lines show $R_{P \rightarrow I}$ and dashed lines show $R_{I \rightarrow I}$. These curves show leakage in the D-plane.}
    \label{fig:power_ratios}
\end{figure}

The most important aspect to consider for this paper, however, is how this cross-polarized beam pattern may contribute to polarization leakage in the context of an intensity mapping interferometer. One can naively try to eliminate polarized emission from one's data by averaging visibilities from perpendicularly polarized feeds: 
\begin{equation}
    V_{I} = \frac{1}{2}(V_{XX} + V_{YY}).
\end{equation}
In this formula, $V$ refers to visibility and $X,Y$ refer to polarizations of the feeds in the baseline. We approximate contributions of unpolarized and polarized emissions to $V_{I}$ in the manner of \citet{Shaw_2015}   
\begin{equation}
    R_{I \rightarrow I} = (E_{x}^{a}E_{x}^{b*} + E_{y}^{a} E_{y}^{b*})P_{ab}^{I}.
\end{equation}
In this formula, $P_{ab}^{I}$ is a matrix selecting for the $I$ Stokes parameter. $E_{x}^{a}$ refers to polarization component $a$ of the electric field pattern of an x-polarized feed. Similarly, the contribution of polarized emission is given by 
\begin{equation}
    R_{P \rightarrow I}^{2} = \sum_{p \in \{Q,U,V\}}|(E_{x}^{a}E_{x}^{b*} + E_{y}^{a} E_{y}^{b*})P_{ab}^{p}|^{2}.
\end{equation}
In Figure \ref{fig:power_ratios}, we present results of leakage calculations for a few frequency channels. These curves are shown in the D-plane and are peak normalized to the maximum value of $R_{I \rightarrow I}$.
In all plots, one finds no polarization leakage along the boresite. 
However, there do exist significant levels off-axis, with the most severe case being $1100$\,MHz with leakage values around 0.4 of the peak copolar value. These levels of polarization leakage are comparable to those estimated for the CHIME telescope in \citet{Shaw_2015}.


\section{Fabrication and Testing of the Feed Antenna}
\label{section: fab}
After completing the design phase of the project, we fabricated a version of the minimum systematics feed. This feed is roughly $0.9$\,m long and $1.2$\,m in diameter at its widest point. Such a large design would not fit in the near-field antenna testing chamber available to us, so a $\frac{4}{13}$ scale version was fabricated instead. In simulations, we found no significant performance difference between the $\frac{4}{13}$ scale design and the original design. 

In this section, the fabrication of this scaled down feed is described. We also include some discussion on what would be done differently for the fabrication of the full-scale feed. Lastly, we  present measurements of the feed's performance and compare these results to simulations.
\subsection{Fabrication of the Scaled Feed}
\begin{figure}
    \centering
    \includegraphics[scale=0.1]{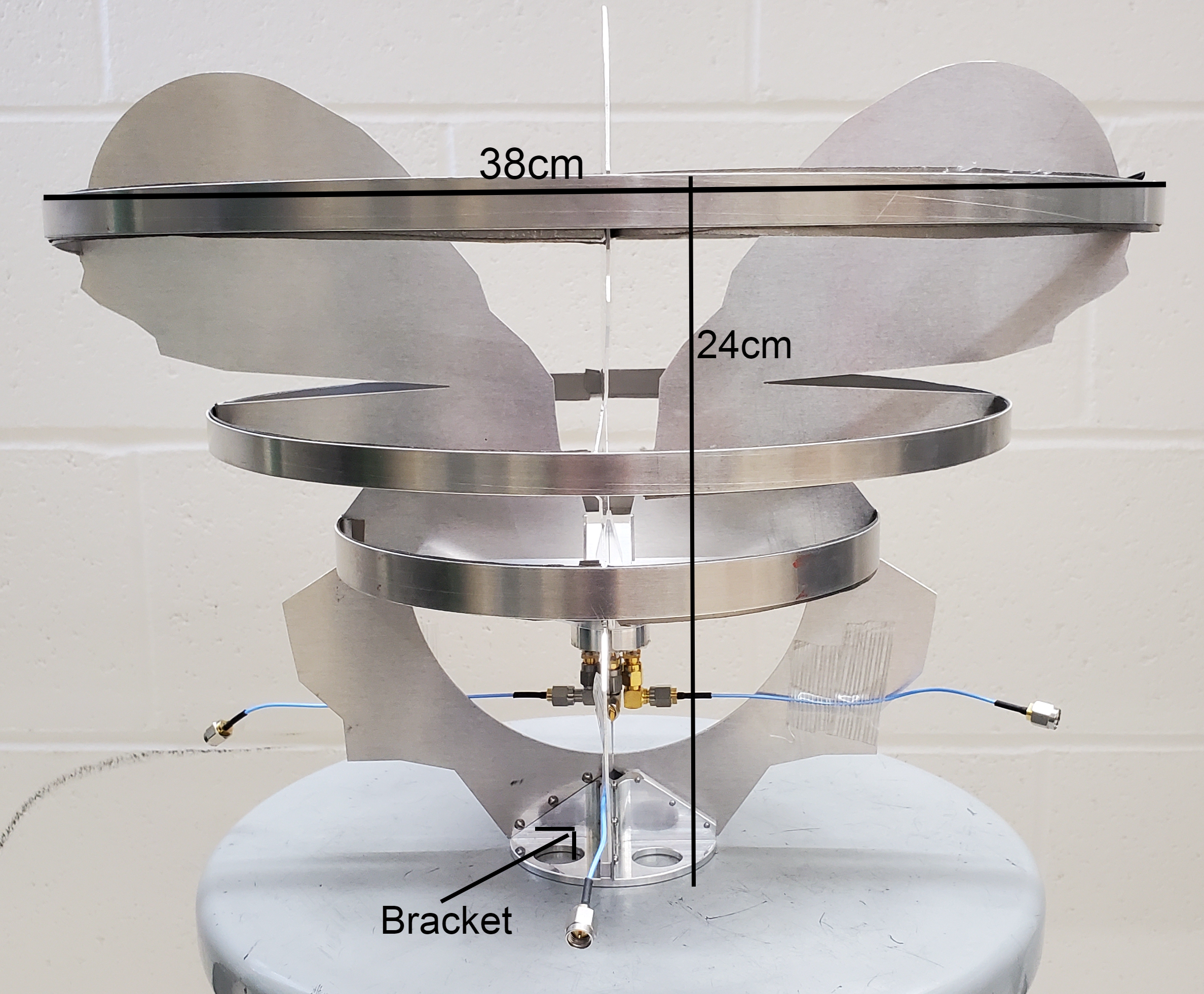}
    \caption{A picture of the $\frac{4}{13}$ scale model of the minimum systematics feed. The balun and coaxial cables are present just for testing. In the future, an impedance matching and amplification module would be used instead.}
    \label{fig:feedfab}
\end{figure}

A photo of the fabricated feed antenna can been seen in Figure \ref{fig:feedfab}. During the fabrication of this feed, a $0.1$\,mm tolerance was requested on all dimensions given. This includes the positions $x_{i},z_{i}$ of points on the profile, differential line pin radius $a_{pin}$, etc.  

The ridges were cut from $1$\,mm sheet aluminum using a laser cutter. Such an approach may work for the full-sized feed, where the ridges are $\approx 1/8"$ thick. However, it may be better to use a water-jet cutter for the full-sized feed.

The cone and differential line shielding were cut from a single block of aluminum. The cone was cut from an aluminum block rather than rolled due to the difficulty of rolling such a thin sheet. Since welding was not an option, the shielding for the differential line was cut from the same block to avoid having to attach the two pieces. Such difficulties would not be present in the fabrication of the full-sized feed. In that case, the use of thicker aluminum would allow for the welding together of pieces and easier rolling of the cone.

An additional aluminum bracket was added to the feed in order to hold the ridges in place. This piece is located on the back of the ridges close to the differential line. It can be seen in Figure \ref{fig:feedfab}. Since it is located in the backlobe of the antenna, this piece makes a negligible difference to the $S_{\rm 11}$ and beam patterns of the feed.

Another aspect to note here is the different absorber used. In the full scale version, Fair-Rite flexible ferrite tile was intended to be used. This material was mentioned in Subsubsection \ref{subsection:rings}. However, data in the frequency range of interest for the scaled down feed could not be found. So, Eccosorb NS-1000 was used instead. In this scaled down design, a layer of absorber $1.2$\,mm thick was used on the largest ring. A $0.6$\,mm thick layer absorber was used on the other two rings.

\subsection{Fabricating a Balun}
\begin{figure}
    \centering
    \includegraphics[scale=0.3]{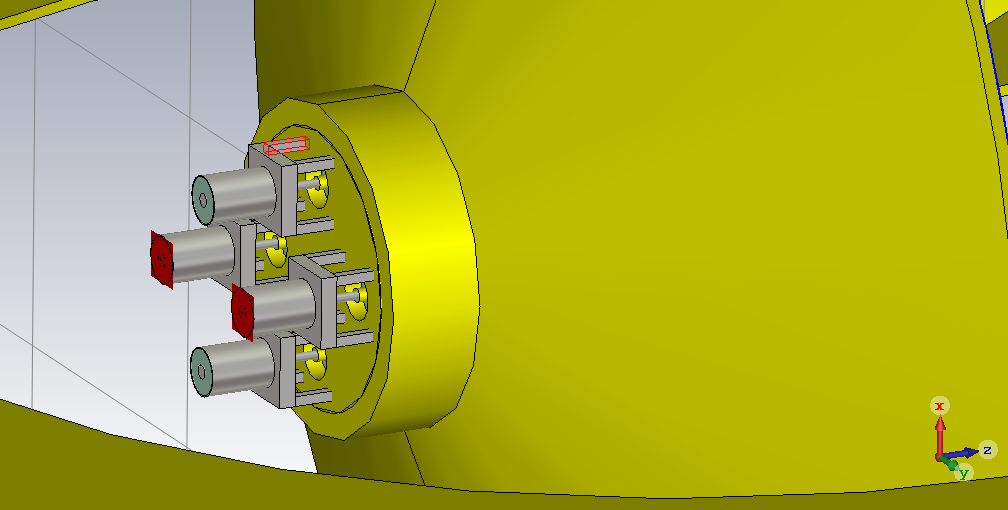}
    \caption{A drawing of the balun used to transition to SMA. These SMA ports then plug into the input ports of a 180 degree hybrid.}
    \label{fig:balun}
\end{figure}

This feed was designed with a differential line as its output. In order to test it, a balun was required to transition from the differential line to SMA connectors. This was done by using a metal disk shown in Figure \ref{fig:balun}. Note that this disk has holes for each pin of the differential line. The diameter of these holes is chosen to form a $50\,\Omega$ transmission line. SMA connectors were then placed on the surface of the disk. The center pins of the SMA connectors touch the pins of the differential line, forming a connection. These SMA connectors are then connected to a $180^\circ$ hybrid, converting the differential signal to an unbalanced one.

This simple balun works, and impacts the beam patterns negligibly. However, it does not provide a particularly good impedance match to the feed. This is acceptable for now, since the main goal was to test whether or not the CST simulations are accurate. In the future an impedance matching circuit similar to HERA's FEM could be used\cite{hera_viv}. 


\subsection{Testing}
In order to make sure our CST simulations were accurate, the $S_{11}$ of the proposed feed was measured using a network analyzer. The beam patterns were then  measured using a StarLab Multiprobe System\footnote{Manufactured by Satimo \url{http://www.mvg-world.com/en/products/field product family/
antenna-measurement-2/starlab}} located in an anechoic chamber at UW-Madison. 

\begin{figure}
    \centering
    \includegraphics[scale=0.5]{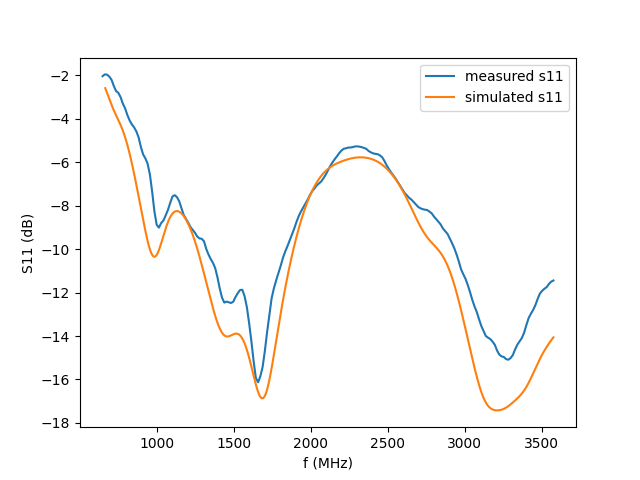}
    \caption{Comparison of the measured $S_{11}$ to the simulated $S_{11}$}
    \label{fig:measS11}
\end{figure}

The resulting $S_{11}$ can be seen in Figure \ref{fig:measS11}. The simulation and measurements here were both done using the $4/13$ scale version of the feed and the simple balun. This $S_{11}$ is reasonably close to the simulated values, being within $1-2\,$dB over the entire band. One will note that these values for the $S_{11}$ are quite high. This is due to the presence of the simple balun in both the simulation and the measurement, which degrades the impedance match. This balun connects a $200\,\Omega$ differential line to a $50\,\Omega$ coaxial line without any attempts at impedance matching. These values would surely improve if one connected the antenna to an output better matched to the differential line. 

\begin{figure}
    \centering
    \includegraphics[scale=0.4]{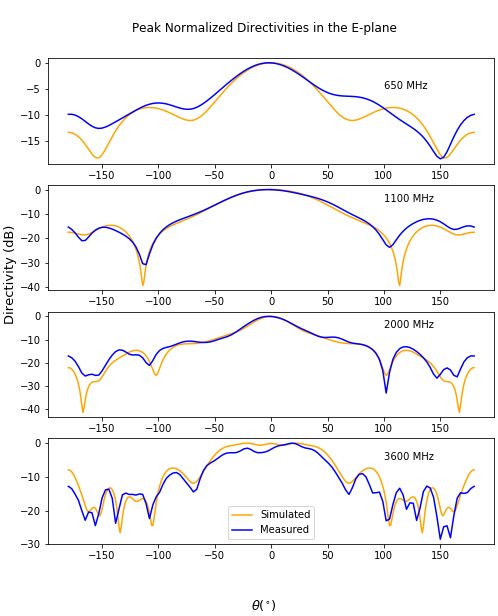}
    \caption{Measurements of E-plane gains for the scaled down feed.}
    \label{fig:Epatts}
\end{figure}

\begin{figure}
    \centering
    \includegraphics[scale=0.4]{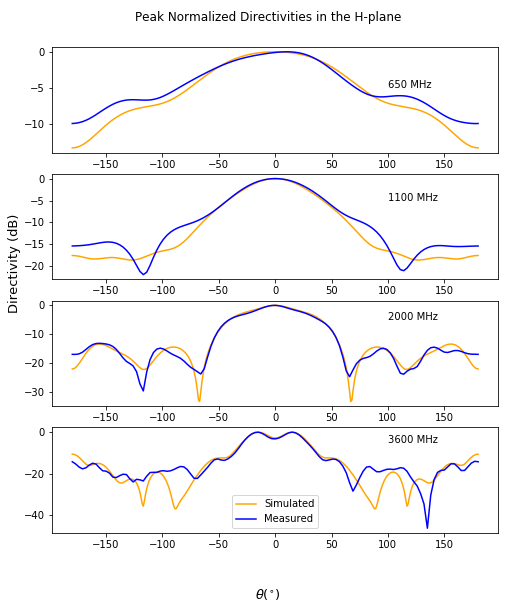}
    \caption{Measurements of H-plane gains for the scaled down feed.}
    \label{fig:Hpatts}
\end{figure}

Some examples of measured beam patterns can be seen in Figure \ref{fig:Epatts} and Figure \ref{fig:Hpatts}. One feature to note is that the patterns are not quite symmetric about $\theta = 0$. This is expected due to fabrication errors and the presence of the $180^\circ$ hybrid inside the anechoic chamber. The only feature that is somewhat concerning is the poorer taper at low frequencies in the H-plane. Having the beam patterns taper off adequately at the rim of the dish is an important feature for reducing mutual coupling. Otherwise, the measured beams appear to be quite close to those produced by our simulations.

\section{Conclusion}
In this paper, we proposed an antenna design optimized for use in PUMA, a proposed HI intensity mapping observatory. The chief performance requirements were to achieve a smooth frequency response, low levels of mutual coupling between antennas in the array, and sensitivity comparable to that of the fiducial PUMA antenna described in \cite{cosmic_visions}.  

An important performance goal was to have the delay-space power kernel of the instrument drop by $50\,$dB as quickly as possible. This goal was set in order to minimize the leakage of Galactic foregrounds beyond the foreground wedge. For the proposed design, it was found that the power kernel dropped to the $-50\,$dB level at a delay corresponding to $k_{||} \approx 0.1\,{\rm hMpc^{-1}}$. This level was set mostly by feed-dish reflections, with the power kernel of the feed alone dropping by $50\,$dB after a delay corresponding to $k_{||} \approx 0.03\,{\rm hMpc^{-1}}$. Reducing these reflections would require using a shallower dish or perhaps an off-axis design. However, these design choices would worsen delay spectrum performance by increasing mutual coupling. Thus, the use of a deep dish with a collar appears to be the best option for maximizing spectral smoothness.

The next performance goal was to minimize mutual coupling. It is important to minimize this coupling due to its effects on the delay spectrum, redundant calibration, and the risk of introducing correlated noise. Based on the simple model shown in \citet{kerns}, the goal was to keep the level of coupling below $-100\,$dB for the entire band. This goal was not achieved, but significant progress was made through optimization. The proposed antenna has coupling at the $-60\,$dB level at low frequencies for closely packed antennas located in the each other's H-plane. Moreover, $-100\,$dB coupling was achieved in the E-plane for frequencies $\nu > 600\,$MHz. 

We also sought to achieve sensitivity close to that provided by PUMA's fiducial antenna parameters. Our proposed feed satisified this goal, giving slightly lower sensitivity at lower redshifts and slightly higher sensitivity at higher redshifts.


The design presented here was optimized with the systematic effects of HI intensity mapping instruments in mind. This design achieved comparable delay spectrum performance to the HERA Vivaldi feed \cite{hera_viv} and provided sensitivity comparable to the fiducial parameters provided for in PUMA telescope proposal. The optimization process for the antenna improved mutual coupling significantly, but the level achieved for the lower end of the band still did not meet the goal. It may be possible to reduce this coupling further using calibration techniques. However, it will likely be necessary to find a way to eliminate this effect in the data analysis stage.

\section*{Acknowledgments}
We thank Trevor Oxholm, Anh Phan, Gage Siebert, David Kwak and Yanlin Wu for feedback on the feed/antenna design. We also thank Doug Dummer and Sara Yaeger of the UW Physics Instrument Shop for fabricating the feed. We thank Chad Seys for computing support. Thanks to Kevin Bandura for helpful feedback on the paper. Thanks to Craig Podczerwinski for fabrication advice. Finally we thank Alex Bouvy, Nader Behdad and Mirhamed Mirmozafari for assistance with the antenna measurements.  This work was partially supported by NSF Award AST-1616554, the University of Wisconsin Graduate School, the Thomas G. Rosenmeyer Cosmology Fund, and by a student award from the Wisconsin Space Grant Consortium.

\appendix{Dimensions of the Ridges}

This appendix gives dimensions of the ridges of the full scale design. If one wishes to make the scaled down version seen in Sec.\ref{section: fab}, then these values can simply be scaled by a factor of $4/13$. First, note that the ridges are made from $1/8$\,'' aluminum plate. For the scaled down design, one would use 1\,mm aluminum plate instead. In Figure \ref{fig:drawing}, one can see that the ridge profile is broken into several sections. The inner edge of the ridge is made up of a series of nodes connected by straight lines. The locations of these nodes are provided in the first set of values in Table \ref{tab:bottomnodes}.

\begin{table}[h]
  \begin{center}
    \caption{Coordinates of all nodes on the ridge shown in Figure \ref{fig:drawing}. The top set of values corresponds to the inner edge of the ridge. The bottom set corresponds to the outer edge of the ridge.}
    \label{tab:bottomnodes}
    \begin{tabular}{l|r} 
      \textbf{z (mm)} & \textbf{x (mm)} \\
      
      \hline
      -54.2 & 27.7\\
      0 & 27.7\\
      41.1 & 38.4\\
      82.4 & 57.8\\
      123.6 & 62.8\\
      164.8 & 64.9\\
      206.0 & 93.8\\
      247.3 & 138.3\\
      288.5 & 151.2\\
      329.7 & 185.4\\
      370.9 & 199.8\\
      412.1 & 235.1\\
      453.3 & 279.1\\
      494.5 & 351.2 \\
      
      \multicolumn{2}{c}{\vspace{1cm}}\\
       \textbf{z (mm)} & \textbf{x (mm)} \\
      \hline
      377.7 & 599.5 \\
      373.3 & 625.5 \\
      334.0 & 625.5 \\
      314.4 & 614.2 \\
      294.7 & 608.6 \\
      275.1 & 614.9 \\
      255.4 & 593.4 \\
      235.8 & 562.3 \\
      216.1 & 531.2 \\
      196.5 & 526.5 \\
      176.8 & 451.8 \\
      157.2 & 171.5 \\
      137.5 & 464.6 \\
      98.2 & 464.6 \\
      78.6 & 370.0 \\
      58.9 & 363.1 \\
      39.3 & 294.7 \\
      0 & 325.0 \\
      -54.2 & 325.0 \\
      -104.2 & 401.1 \\
      -154.2 & 393.8 \\
      -204.2 & 385.0 \\
      -254.2 & 333.7 \\ 
      -308.8 & 330.2 \\
      -308.7 & 264.2 \\
      -322.7 & 198.1 \\
      -414.6 & 132.1 \\
      -322.7 & 198.1 \\
      -414.6 & 132.1 \\
      -322.8 & 0 \\ 
      -304.2 & 0 \\ 
      
    \end{tabular}
  \end{center}
\end{table}

Then, there is a circular section connecting the inner edge of the ridge to the outer edge. This circular section has a radius of $168$\,mm. The outer edge is also made up of nodes connected by straight lines. The node positions of this edge are provided in the second set of values in Table \ref{tab:bottomnodes}.
      

Lastly, a quarter circle is used for the backshort section of the ridge. This quarter circular has a radius of $250$\,mm.





%

\clearpage
\bibliographystyle{ws-jai}
\bibliography{bibliography}
\end{document}